\documentclass[CRPHYS,Unicode,manuscript]{cedram}

 \usepackage[all]{xypic}
\usepackage{mathtools}

\DeclarePairedDelimiter\floor{\lfloor}{\rfloor}
\title{Fluctuations in crystalline plasticity }

\author{\firstname{Jérôme} \lastname{Weiss}\IsCorresp}
\address{ISTerre, CNRS/Université Grenoble-Alpes, 38041 Grenoble, France}
\email[J.Weiss]{jerome.weiss@univ-grenoble-alpes.fr}
\author{\firstname{Peng} \lastname{Zhang}}
\address{State Key Laboratory for Mechanical Behavior of Materials, Xi'an Jiaotong University, Xi'an, 710049, China}
\email[P. Zhang]{zhangpeng.mse@xjtu.edu.cn}
\author{\firstname{O\u{g}uz Umut} \lastname{Salman}}
\address{CNRS, LSPM UPR3407, Sorbonne Université Paris Nord, 93430, Villetaneuse, France}
\email[O.U.Salman]{umut.salman@lspm.cnrs.fr}
\author{\firstname{Gang} \lastname{Liu}}
\address{State Key Laboratory for Mechanical Behavior of Materials, Xi'an Jiaotong University, Xi'an, 710049, China}
\email[G. Liu]{lgsammer@xjtu.edu.cn}
\author{\firstname{Lev} \lastname{Truskinovsky}}
\address{PMMH, CNRS UMR 7636, ESPCI ParsiTech, 10 Rue Vauquelin, 75005, Paris, France}
\email[L.Truskinovsky]{lev.truskinovsky@espci.fr}
\thanks{The authors acknowledge the support of the French-Chinese ANR-NSFC grant SUMMIT (ANR-19-CE08-0010-01 and 51761135031). P.Z. acknowledges additional support from the China Scholarship Council and China Postdoctoral Science Foundation, grant 2019M653595} 

\begin{abstract} 
Recently acoustic signature of dislocation avalanches in HCP materials was found to be long tailed in size and energy, suggesting critical dynamics.  Even more recently,  the   intermittent plastic response  was found to be generic   for micro- and nano-sized systems independently of their crystallographic symmetry. These rather remarkable discoveries are  reviewed  in this paper  in the perspective of  the  recent  studies performed in our group. We discuss  the physical origin and the scaling properties of plastic fluctuations and address the  nature  of their dependence on  crystalline symmetry, system size, and disorder content. A particular emphasis is placed on the   associated emergent behaviors,  including the formation of dislocation structures, and on our ability  to  temper plastic fluctuations  by alloying.  We also discuss  the  "smaller is wilder" size  effect that culminates in a paradoxical  crack-free brittle  behavior  of  very small, initially dislocation free crystals. We show that the  implied transition between different rheological behaviors  is regulated by the  ratio of length scales $R=L/l$, where  $L$ is the system size  and $l$  is the   internal length. We link this new  size effect with other related phenomena  like size dependence of  strength ("smaller is stronger") and the size induced switch between different  hardening mechanisms. One of the technological challenges in nanoscience is to tame  the intermittency of plastic flow.    We show that this task can be accomplished by generating   tailored quenched disorder which allows one to control micro- and nano-scale forming and opens new perspectives  in micro-metallurgy and structural  engineering of ultra-small load-carrying elements. These results could not be achieved by conventional methods that do not explicitly consider the stochastic nature of collective dislocation dynamics.    
\end{abstract}

\begin{document}

% Use the \maketitle command after the abstract
\maketitle

%% Beginning of text
%% Abridged versions
%% 1. English one if the paper is in French
% \selectlanguage{english}
% \section*{Abridged English version}
% <your text here>
% \selectlanguage{french}
% to go back to main language.

%% 2. French one if the paper is in English
 \selectlanguage{french}
 \section*{Version française abrégée}
Malgré une étude précurseur de Becker et Orowan en 1932 sur le Zinc, l'analyse des fluctuations dans la dynamique de la déformation plastique des matériaux cristallins a été pendant longtemps négligée, probablement du fait que dans la plupart des matériaux métalliques d’intérêt industriel ces fluctuations sont indétectables en termes de comportement mécanique aux échelles macroscopiques. La situation a changé drastiquement il y a une vingtaine d’années lorsque, d’une part, l’enregistrement des signatures acoustiques des avalanches de dislocations dans certains matériaux hexagonaux a montré que ces dernières pouvaient être distribuées en loi de puissance, suggérant une dynamique critique, et d’autre part il a été observé que la plasticité des systèmes de taille micro- et nano-métrique devenait intermittente pour la plupart des matériaux non-alliés. Dans cet article, nous discutons, sur la base de récents travaux et dans le cadre de la physique statistique, de la nature et des propriétés statistiques et d’échelle de ces fluctuations plastiques en fonction de la symétrie cristalline, de la taille du système considéré, ainsi que du désordre interne, qu’il soit émergent (structures de dislocations) ou ajusté par des techniques d’alliage. On met ainsi en lumière des effets de taille très prononcés sur la stochasticité de la déformation plastique, un rapport d’échelle $R=L/l$ entre la taille finie du système $L$ et une échelle interne $l$ jouant un rôle majeur pour expliquer ces transitions de comportement.  On discute également le lien avec d’autres effets d’échelle, sur le seuil d’écoulement plastique ou la nature des mécanismes de durcissement, et on montre comment les techniques d’alliage peuvent réduire ces instabilités plastiques.  Ceci ouvre la voie vers une métallurgie et des pratiques d’ingénierie aux échelles sous-microniques tenant compte du caractère stochastique intrinsèque de la plasticité à ces échelles, et tentant de le supprimer ou de l'atténuer.
\selectlanguage{english}
%to go back to main language.

% Example of section
\section{Introduction}

Beyond the qualitative study  of  Becker and Orowan (1932), the fluctuating nature of crystal plasticity has been largely  overlooked in the XX$^{th}$ century, not in the least, because for most materials of engineering interest  these fluctuations remained almost  undetectable and considered as negligible  comparing to macroscopic response  at bulk scales. The situation  changed dramatically at the turn of this century when new technology opened a  way to  precision measurement of  acoustic emissions accompanying plastic deformation. First, the acoustic signature of dislocation avalanches in HCP materials was found to be long tailed in size and energy, suggesting critical dynamics.  Second,  the   intermittent plastic response  was found to be generic   for micro- and nano-sized systems independently of their crystallographic symmetry. 

% Example of subsection
\subsection{Continuum mechanics vs discrete approaches}

Since the "miraculous year",   1934, lattice dislocations are universally accepted as the main carriers of plastic deformation in crystalline solids \cite{orowan1934,orowan1934b, orowan1934c, polanyi1934, taylor1934}. The proposed dislocation theory was intrinsically  discrete: plastic deformation results from the 'quantized'   activity of  topological defects. An alternative singular   concept of dislocations   was developed much earlier, at the beginning of the XXth century, and it operated  within a continuum framework of solid mechanics \cite{Timpe1905, Volterra1907, hirth1985}. The singularity centered understanding of dislocations  was not undermined by the inherently discrete theory, for instance, the continuum theory of dislocations   is still actively used  in seismology \cite{savage1980}. It is now universally accepted that the type of description   should depend on the scale of coarse-graining. In particular, it has been rigorously shown  that the  'quantized' lattice dislocations take the form of the associated elastic singularities when the Burgers length scale is small comparing to other length scales in the problem.  

The subsequent  progress in  crystalline plasticity was based  on the advances in  our understanding  of the structure of  the cores  of  individual topological defects, in quantifying   their short range interaction, and more recently in finding correlations in their collective dynamics.  The  continuum mechanical approaches to  dislocation-driven plasticity were developed in parallel to lattice-based theories, starting  from  the pioneering works of Nye \cite{nye1953} and Kröner \cite{kroner1959} who built the foundations of a field theory linking  continuum  displacement field  to  dislocation density tensor.  For recent  developments along these lines  see  \cite{hutchinson1997,acharya2000,acharya2001,fressengeas2011,valdenaire2016}. 

Comparing to meso-scale modelling approaches, explicitly accounting for  individual dislocations (or dislocation segments) and their interactions \cite{kubin1992,kubin2013}, and to fully microscopic  molecular dynamics simulations \cite{rodney2004,zepeda2017}, the  conventional phenomenological continuum models of crystal plasticity offer  the most rough  description in the sense that they   capture only the main effects of plastic yielding. They achieve the goal of representing complex geometries and loading conditions by drastically reducing the number of degrees of freedom.  The implied spatial coarse-graining and temporal averaging  should be based on the  identification  of  small internal length and time scales and should rely on  the  formal homogenization techniques allowing one to  link the   continuum variables   with their  lattice counterparts. 

However, such a program has never been implemented rigorously and the existing continuum approaches to crystal plasticity  are largely phenomenological.  Already the very idea of a pure continuum theory  has   several fundamental shortcomings. In particular:  

(i) Identifying of the appropriate internal length  scale \textit{l} remains  an unsolved problem. In metals with multiple slip systems, such as FCC, it is known  that dislocation patterns emerge spontaneously \cite{madec2002,kubin2002}. These patterns may or may not  be associated with a well-defined characteristic scale $l_p$ \cite{hahner1998}. In cases they do, $l_p$ may represent    a typical cell size or  a spacing between persistent slip bands\cite{laird1986} and  is generally of the order of a few µm in FCC materials.  It would then be    inversely proportional to the yield stress due to the similitude relation (see more below) \cite{sauzay2011,kubin2013}. When such a scale can be unambiguously defined, it may be tempting to identify  the  homogenization scale $l$ with $l_p$. With all this said, the determination of $l_p$, as well as the check of the correctness of the similitude relation, remain fully empirical. Moreover, in crystalline structures characterized by a strong plastic anisotropy, such as HCP crystals, the implied  dislocation patterns with a well-defined characteristic size do not form at all \cite{weiss2019}, which  leaves the question of the averaging scale completely open. 

(ii) There has been an increasing evidence of the existence of temporal  fluctuations  in crystalline plasticity induced by cooperative motions of dislocations. The  intermittent dynamics was qualitatively recognized   in HCP crystals long ago \cite{becker1932}, however, it has not been  analyzed quantitatively till the turn of this century \cite{weiss1997,miguel2001}. This became a bigger  issue when it was realized that, upon decreasing the system size below few µm - noticeably, the same order of magnitude as the  characteristic scale $l_p$ -, plastic deformation becomes jerky independently of  crystal symmetry \cite{dimiduk2006,brinckmann2008,ng2008}. It  raised major concerns regarding  manufacturing, and reliable utilization of ultra-small machinery \cite{csikor2007,hu2016}. By construction, the phenomenological  continuum models and field theories are unable to deal with such   fluctuations, unless stochastic closures  are implemented implicitly \cite{zaiser2005,zaiser2013}. 

(iii) by construction, the continuum dislocation density tensor only accounts for  dislocations generating a net plastic distortion at a given    scale \textit{l}, the so-called "geometrically necessary dislocations (GND)".  It, therefore,    ignores the "statistically stored" dislocations (SSD) with zero net Burgers vector \cite{hirth1985}. This is a strong limitation as SSD are  tightly linked to hardening. In principle, this problem can be "resolved" through a decomposition of the dislocation density tensor into contributions from different slip systems combined with a very high spatial resolution, well below the average dislocation mean-free path \cite{xia2015}. At that  resolution  any dislocation becomes geometrically necessary, however, it would  completely eliminate the advantage of a continuum description \cite{monavari2018}. Instead, the  continuum descriptions  fully focused on the internal elastic fields associated with GND  can hardly deal with dislocation nucleation  and other  processes involving short-range interactions such as multiplication, annihilation, or the formation of junctions. There are however, recent developments in this direction that are promising   \cite{monavari2018}. 
  
Note that some macroscopic plastic instabilities and jerky plastic flow  in particular alloys have been linked to  dynamic strain aging due to the diffusion of solutes towards dislocation cores and the unpinning of dislocations from these point defects \cite{lebyodkin1995,ananthakrishna1999,bharathi2002}. This interesting but somewhat special problem,  related to the emergence of  non-intermittent fluctuations,  can be  analyzed in  a classical continuum framework.  We leave  such phenomena   outside   the scope of the present paper  focusing   on pure crystals, or systems containing passive quenched disorder.  Systems with intermittent fluctuations  have been so far beyond reach for continuum theories.

\subsection{Thermodynamic approach}

Unlike elasticity, plasticity is a dissipative process generating  heat   and   irreversiblity.   This implies that a conservative  (reversible) purely mechanical description is inadequate, and argues instead for a thermodynamical approach. However, it is immediately clear that weakly nonequilibrium thermodynamics,  which has been successful in describing linear viscoelasticity and heat conduction,  would not be sufficient because plasticity is not linearizable.  Since the adequate strongly nonequilibrium thermodynamics  does not exist, the possibility to model plasticity directly in statistical mechanics terms by viewing crystal as an ensemble of interacting  dislocations   was already envisaged in the early days of dislocation theory \cite{berdichevsky2019}. 

The elusive character of this initial hope was however, soon realized and for multiple reasons. If point interactions within  a gas are short-ranged (collisions), dislocations, represented by line segments,  interact   both  at short-range (direct entanglements and depinning) and at long range (transmitted by elastic fields). The combination of  threshold type interaction and the  extended nature of the interacting defects gives rise in a driven system to a rich    dynamical behavior  characterized by cooperative effects and  self-organization.  The emerging  scaling laws is thus just a signature of the underlying highly correlated collective dynamics. Since  dislocations are defects with such  high energy that ordinary thermal fluctuations cannot affect,  the system remains far from thermal  equilibrium \cite{kubin2013}. Instead,  complex metastable  out-of-equilibrium dislocation patterns emerge and intermittently store and release the energy provided by the loading device. The energy   effectively flows through the system arriving at macro scales and dissipating  at the scale of dislocation cores.
The emerging downscale energy cascade is probably as complex as in the case of turbulent flow with scaling similarly setting in the intermediate 'inertial' range.

Despite these impediments, there have been  recent  attempts to build a thermodynamic theory of crystalline plasticity intending to rationalize the phenomenology of dislocation patterning and   understanding of  strain-hardening from first principles \cite{berdichevsky2019,langer2010,langer2020}. Such theories have been shown successful   at elevated  temperatures and  nonzero  loading rates  in crystals with significantly inhibited long range dislocation interactions. In such systems, the role of elasticity is weak, and the self-organization of defects is repressed, which allows at least partial thermodynamic equilibrium to be reached. However, despite these successful attempts, it is probably safe to say that a broadly applicable thermodynamic theory of plasticity remains an illusion. 

\subsection{The goals of this review}

Our  aim here is  to discuss  in some length recent advances in the characterization and modelling of plastic fluctuations in   space, time and energy domains. We place particular emphasis on the fact   that  plastic  flow in crystals  can occur through intermittent slip avalanches that are  power-law distributed, i.e. in a scale-free manner \cite{weiss1997,miguel2001}.    While the presence  of such  highly correlated fluctuations clearly suggests    the \textit{critical} character of collective dislocation dynamics,  the  nature of the underlying  self organization mechanism  is still highly debated \cite{friedman2012,ispanovity2014,song2019}.   The literature devoted to  plastic intermittency, dislocation avalanches, and the associated  criticality  is steadily growing and the  exhaustive review of the subject is far beyond the scope of this paper. The early work was reviewed in \cite{zaiser2006} and more recent extended discussions can be found in \cite{uchic2009,greer2011, papanikolaou2017, maass2018,cui2018,sethna2017,ovid2018,dehm2018,groma2019}. Instead our goal is to focus on a few  subjects that some of us have actively taken  part in exploring. 

Mainly, our discussion is  centered around  the physical origin of the  scaling  behavior.  Along the way, we also address the  nature  of the dependence of the critical exponents   on  crystalline symmetry, system size, and disorder content. We only touch upon such important emergent behaviors as  the formation of dislocation structures and system size events.

The recent discovery that the scaling in crystal plasticity   is probably not universal,  depending  on crystal symmetry \cite{weiss2015,weiss2019}, system size \cite{papanikolaou2012,zhang2017}, quenched  disorder \cite{zhang2017}, and the presence of hardening \cite{zaiser2007,weiss2019b}, reveals  rich and  complex physics of the  underlying self-organization process  that still remain largely unexplored \cite{zhang2020}. Without attempting to make a comprehensive review of the implied elemental  mechanisms,  we concentrate   on the  recent work addressing: (i) experimental tracking of plastic fluctuations either directly from stress-strain records  \cite{zhang2017,zhang2020b}, or from acoustic emission (AE) in bulk materials \cite{weiss2015,weiss2019b,LHote2019}; (ii) numerical simulations based on  the  minimal automaton model of plastic flow in crystals proposed in \cite{salman2011,zhang2020}; (iii) stochastic modeling of  mesoscale  crystal plasticity introducing rheological closure relations with multiplicative noise \cite{weiss2015}.  Our various  comments on the recent progress in the field  are therefore presented in the perspective of the  studies performed in our group.

 We  begin by  summarizing the current understanding of the nature of scaling in pure materials (not alloyed).  In particular we explain  why the associated exponents  depend on crystal symmetry and system size. We  then discuss how the transition   from mild (Gaussian-like) to wild (power law distributed ) fluctuations  depends on an internal length scale $l$ and use the obtained understanding to build a  connection with plastic hardening.  Our next subject is   the  "smaller is wilder" size  effect that culminates in a paradoxical  crack-free brittle  behavior  of  very small, initially dislocation free crystals. We  show that such  transition between different rheological behaviors  is regulated by the  ratio of length scales $R=L/l$, where  $L$ is the system size. We link this relatively new  size effect with some other well studied  phenomena  like the size dependence of  strength ("smaller is stronger") and the size induced switch between different  hardening mechanisms.  
 
 After gaining the insight regarding   the fluctuational response of pure crystals, we   broaden the scope of our discussion to include alloys, and show that in submicron samples  altering the alloying-induced quenched   disorder may model variation of the system size.  We show that these theoretical considerations can be   rationalized within a numerical model and obtain quantitative relations between the system size and the scaling exponents.  
 
 One of the technological challenges in nanoscience is to tame  the intermittency of plastic flow.    We show that this task can be accomplished by generating   tailored quenched disorder which allows one to control micro- and nano-scale forming.  This recent discovery  opens new perspectives  in nanometallurgy aiming not only at improving mean properties (e.g. strength) but also at tempering associated deleterious fluctuations and variability. 
 
 To conclude we stress once again  that the problems, posed by the emerging science of structural  engineering of sub-micron load-carrying elements, cannot be solved by using  conventional methods of continuum  plasticity. New methods are needed,   taking explicitly into account the stochastic nature of collective dislocation dynamics, and this review should be viewed as a small step in this important scientific direction.

\section{Pure  materials }

\subsection{Historical background}

The first reported evidence of   intermittent plasticity dates back to the pioneering study of Becker and Orowan, who observed during the deformation of Zinc crystal rods  a succession of sudden strain jumps of vastly different sizes \cite{becker1932}. It is worth stressing that these experimental results likely played a substantial role in elaborating dislocation theory  by Orowan two years later \cite{maass2018,orowan1934}.  Forty years after Becker and Orowan, plastic bursts were observed again in  Zinc crystals, still directly on stress-strain curves  \cite{tinder1973}. As we are going to explain below, it is not incidental that all these initial  obsevations of jerky plastic flow were made on HCP crystals  characterized by a strong plastic anisotropy.

Outside these pioneering observations, the intermittency of plastic deformation was only episodically addressed during the XX$^{th}$ century. In classical continuum theory, it was implicitly assumed  that the inevitable fluctuations resulting from discrete  dislocation motions average out and become invisible at  engineering scales.  These fluctuations were believed to be associated with a particular mesoscopic scale that is much smaller than the scale of the observations. They were also perceived as  roughly  Gaussian  or  \textit{mild} in the terminology of Mandelbrot. 

However, experimental   investigations of plastic fluctuations  continued to challenge this picture. In  the late 60's, a new generation of experiments was performed  using indirect monitoring tools  which   switched attention from stress-strain curves to  acoustic emission (AE) accompanying plastic flow. Several authors reported acoustic bursts during  plastic deformation in  materials with different crystal symmetries, and interpreted them   as a signature of slip events resulting from cooperative dislocation motions \cite{fisher1967,james1971}. In FCC metals, such burst-like AE activity was scarce, with maximum intensity at plastic yield but fading away rapidly as the material strain-hardened.  It was also superimposed on the  so-called continuous AE \cite{imanaka1973,kiesewetter1976}. This slow evolving    background 'noise' was interpreted as resulting from the cumulative effect of numerous, small and \textit{uncorrelated} dislocation motions, in full contrast with the cooperative motions giving rise to discrete AE bursts. The AE source model that was used to link the characteristics of the recorded AE signals to local dislocation motions is  detailed elsewhere \cite{rouby1983,weiss2015}. Very generally,  this model assumes that for discrete AE signal, the radiated acoustic energy $E_{AE}$ integrated over the waveform duration is proportional to the mechanical energy dissipated by the plastic event at the source. For continuous AE, the acoustic power $dE_{AE}/dt$ is taken as a proxy of the plastic strain-rate. 

Now, nearly 50 years after these historical studies, it is rather striking to discover how  many  modern  issues related to  plastic fluctuations have been already implicitly identified  at that time.
Those include  the role of crystal symmetry, the influence of   strain hardening, and  the possible coexistence of large cooperative motions (dislocations avalanches) with small uncorrelated motions.  However, these early studies neither provided the   quantitative  analysis of    fluctuations   nor  did they offer   a framework for the  interpretation of the observed intermittency. 

A few decades later; the whole subject  was reactivated  when AE was finally carefully recorded during the plastic deformation of ice single crystals, and the obtained time series were statistically  post-processed \cite{weiss1997,miguel2001}. Subjected to uniaxial compression, this  HCP material with a particularly strong plastic anisotropy, unexpectedly revealed a fully intermittent AE signal characterized by a succession of   bursts that were  power-law distributed in energies. The obtained data could be  fit using  a relation for the probability density  $P(E_{AE}) \sim E_{AE}^{-\kappa_E}$.  The value of the   exponent $\kappa_E$ was found to be somewhat close to the mean-field value 1.5 \cite{friedman2012,uhl2015,salje2014} and independent of either the  applied stress or the  temperature \cite{weiss1997,miguel2001}.

Using  the same material, it was further shown  that the  dislocation avalanches are clustered in time. More precisely, it was demonstrated that the larger the energy of an avalanche, the larger, in average, is the avalanche occurrence frequency immediately after that avalanche.  These observations  were interpreted as an evidence of  'aftershocks' triggered by the main    stress redistribution \cite{weiss2004}. Spatial clustering, i.e. a fractal structure of avalanche distribution, was detected as well. Moreover, spatial and temporal distributions of avalanches   were found to be  linked.  More precisely,  aftershocks turned out  to be triggered in the vicinity of their 'mainshock', which is a  consequence of stress transfer decay with the distance from the mother avalanche \cite{weiss2003}. All these observations were later confirmed on other HCP materials such as Zinc or Cadmium, with the  exponent $\kappa_E$ taking value close to that observed for ice \cite{richeton2006}. 

The obtained results  argue that in HCP crystals the collective dynamics of dislocations  self organizes towards criticality.  Consistently   with earlier  observations of jerky stress-strain curves  \cite{becker1932,tinder1973}, plastic deformation in such materials takes the form  of well separated avalanches. The    statistics of these avalanches was found to be  of power law type and their  maximum size was  constrained only by the size of the system.  In Mandelbrot's terms this mean that the   fluctuations are \textit{wild}  suggesting  that quasi-statically  driven HCP crystals are inherently critical. 

Intermittent, power-law distributed plastic fluctuations were also identified from both AE measurements and high-resolution extensometry in a FCC material, copper, at least in the early stages of deformation \cite{weiss2007}. However, compared to avalanches in HCP materials, the  detected plastic bursts in copper were much more sporadic, which confirmed  observations made in earlier   studies \cite{imanaka1973,kiesewetter1976}. The scarcity of avalanches could not be truly quantified  at that time and its  conceptual importance  was understood only recently  \cite{weiss2015}, see more about this below. 
 
Overall, the  AE measurements in both HCP and FCC  crystals  appear to be  arguing for a scale-free, critical character of crystalline plasticity.  We recall that this is  at odds with the classical paradigm of dislocation-mediated plasticity as a smooth flow resulting from small, uncorrelated and fundamentally similar increments and that was also the picture emerging from the observations on bulk BCC materials. All this naturally raises the question whether  these two conflicting pictures can be  reconciled. A related question is whether  the critical nature of crystalline plasticity, at least  in HCP and FCC crystals,  is   universal and  is characterized by a set of critical exponents that are independent of the particular material and its purity, of the  size and shape of the sample and of the loading conditions.

In what follows, we try to answer these questions addressing separately  the role of crystal symmetry, the system size $L$ and  the quenched microstructural disorder.  We will not address  the role of grain boundaries (GB) because this question has received so far only very limited attention. There are, however,  experimental and numerical studies showing that GBs can hinder the development of  dislocation avalanches, although a spatially correlated plastic activity can spread in polycrystals  over much larger distances \cite{weiss2019,richeton2005,niiyama2016,richeton2005b}. 

\subsection{From  subcritical to  supercritical plasticity}

Over the last years, the growing interest towards manufacturing devices at micro- to nano-scales challenged the classical, size-independent approaches of material engineering and called for   new approaches to mechanical characterization of materials at such small sizes \cite{zaiser2006,greer2011}. Compression tests from a flatbed indenter on µm to sub-µm pillars became a standard tool for such characterization \cite{uchic2004,dimiduk2006,uchic2009}. Other methodologies have been proposed to explore even smaller (few tens of nm) systems, such as tensile tests on nanowires \cite{richter2009} or compression of nanoparticles \cite{sharma2018,mordehai2018}. 

The first outcome of these studies was a discovery of a dramatic size effect on strength $\tau_y$   which is usually defined as the shear stress reached at some arbitrarily specified amount of plastic deformation  \cite{uchic2004}. This "smaller is stronger" phenomenon is  usually expressed through a scaling $\tau_y \sim L^{-\alpha}$, with $\alpha$ an empirical exponent varying for metals in the range 0.2-1  \cite{greer2011,dunstan2013,schneider2009}. The conventional  interpretation of this size effect is  the increased role of free surfaces in small systems. The associated mechanisms are known as   source truncation \cite{parthasarathy2007} or  dislocation  starvation \cite{greer2006}. In initially dislocation-free nanoparticles, yield stresses close to the theoretical shear strength have been reported and linked to a size specific particular dislocation nucleation mechanism \cite{sharma2018}. 

Besides the size effect on strength, compression tests on micropillars of FCC \cite{dimiduk2006} and BCC materials \cite{brinckmann2008,zaiser2008}  revealed the stress-strain curves with an anomalous presence  of intermittent strain bursts. These bursts   were shown to be power-law distributed with the probability distribution  $P(s) \sim s^{-\kappa}$ where $s$ is the burst  size.  

This observation was  in contrast with the scarcity of such plastic bursts at bulk scales, and the association  of  these materials with a smooth macroscopic response, a dominance of multislip and a presence of a single scale in dislocation patterning. It suggests that there is 
a  distinct fluctuation related size effect  which was  coined as "smaller is wilder" \cite{weiss2015}. A wealth of experimental work  confirms the emergence of  intermittent plastic bursts in sufficiently small crystals of both  pure materials and alloys. The ubiquity of power-law distributed plastic bursts at small  $L$ triggered an intense debate regarding  the physical nature of this phenomenon \cite{friedman2012,ispanovity2014,papanikolaou2012,lehtinen2016}. 

In this section we consider pure materials with different crystalline structures with the goal of  exhibiting the  rich landscape of plastic fluctuations and showing how it evolves with the system size.  The effect of extrinsic disorder (alloying)  will be studied in following sections.

\begin{figure}[h]
 \includegraphics[width=0.8\linewidth]{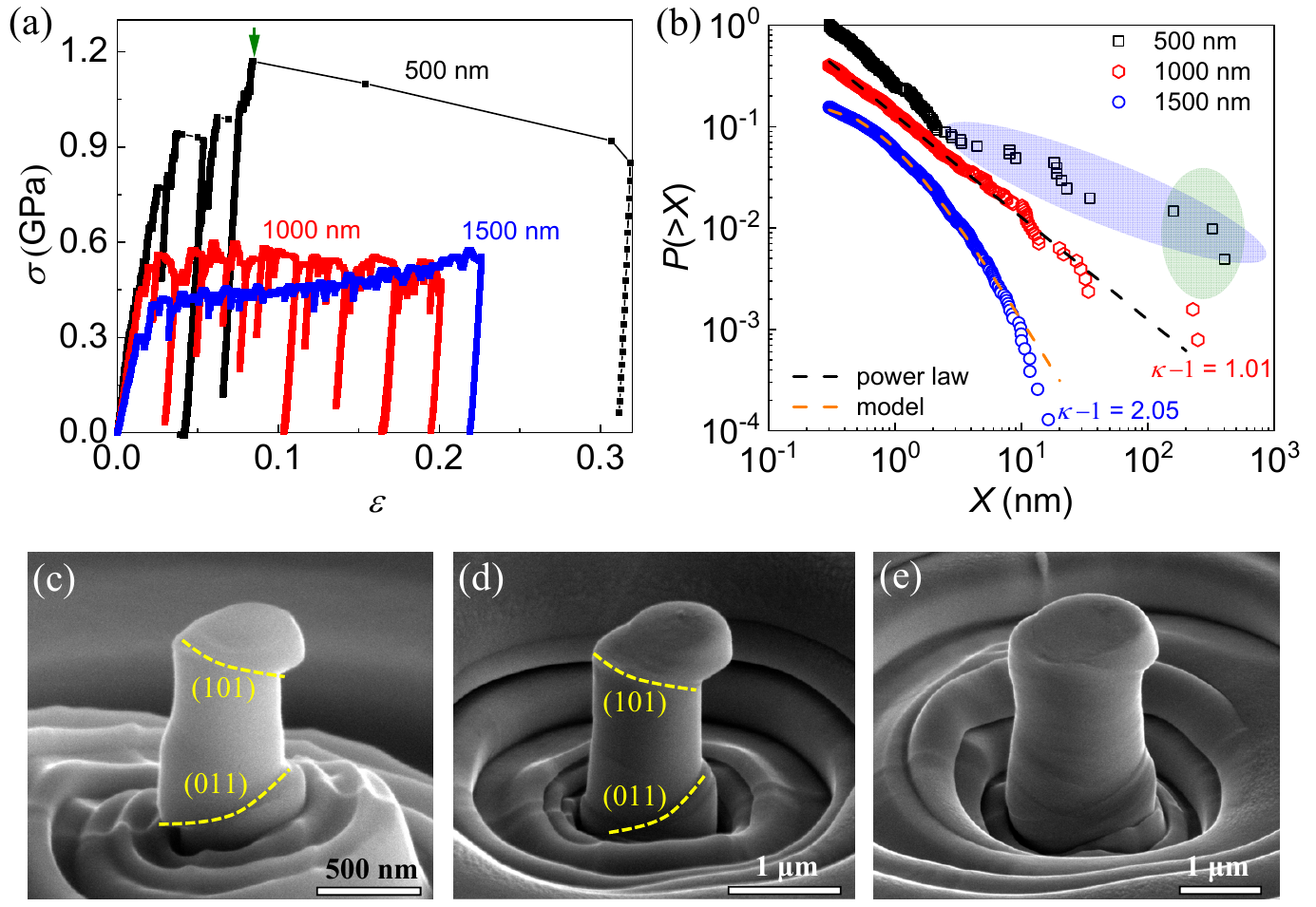}
 \caption{From mild to supercritical fluctuations in Mo micropillars. (a) Stress-strain curves (shear). (b) Cumulative distributions of plastic displacements $X$ detected over the entire loading. The events marked in violet correspond to supercritical events, while those marked in green are dragon-kings, system-spanning events. Dashed line represent the fits of the data with eq. (2). (c) to (e) SEM images of (c) 500 nm, (d) 1000 nm and (e) 1500 nm [112]-oriented micropillars after compression. The marked slip traces indicate the locally single-slip nature of the plastic flow in 500 nm and 1000 nm pillars. Adapted from \cite{zhang2020}. }
 \label{fig:mildtosup}
 \end{figure}

Figure \ref{fig:mildtosup} shows experimental results: in (a) the true stress - true strain relations, and in (b) the cumulative distributions of  displacement burst sizes $X$ (in nm).  In  (c) we illustrate the deformation morphology for compression tests for  µm to sub-µm pillars of a high-purity BCC material, Mo \cite{zhang2020}. In these experiments, the loading configuration was designed to ensure that twinning was absent, i.e. plasticity was accommodated by dislocations only. A size effect on the yield stress is apparent on Fig.\ref{fig:mildtosup}(a). The plastic displacement jumps along the compression axis were extracted from the force-displacement raw data using a methodology  detailed in \cite{zhang2017}. An avalanche was defined as a plastic process characterized by a dissipation rate much greater than the imposed loading rate. The size $s$ of the dislocation avalanche can be linked to the plastic jump $X$, which is assumed  to scale with the cumulative distance covered by all mobile dislocations during the avalanche \cite{maass2013}. The distributions presented on Fig. \ref{fig:mildtosup}(b) contain results from at least four samples of the same size $L$. 
 
 Figure \ref{fig:mildtosup}(b) illustrates the  "smaller is wilder" size effect. For the larger pillars ($L$=1500 nm), the relatively smooth mechanical response (Fig.\ref{fig:mildtosup}(a)) produces  a distribution of plastic fluctuations which clearly has nothing to do with  power law, instead revealing a characteristic size $X_0$ (see more below) and corresponding to a \textit{sub-critical} regime. At an intermediate sample size $L$=1000 nm, the nature of plastic fluctuations changes drastically, with a power law distribution of avalanche sizes, $P(>X) \sim X^{-(\kappa-1)}$ emerging over three orders of magnitude, without detectable lower or upper cut-offs, and with $\kappa \simeq 2.0$. A robust maximum likelihood methodology has been used in all our analyses to estimate such exponents as well as to detect possible lower or upper cut-offs \cite{clauset2009}. The  power law statistics in this range of crystal sizes characterizes a scale-free, critical dynamics, reminiscent of what was observed in several other studies (e.g. \cite{miguel2001,zaiser2006,friedman2012}).  
 
As we  decrease further  the system size to 500 nm, the nature of plastic fluctuations changes  again, producing a regime characterized by the presence of  large scale outliers coexisting with a power-law range at small scales. The appearance of "dragon-kings" can be interpreted as a transition towards a \textit{supercritical} dynamics \cite{sornette2012}. The largest of the outliers  correspond to system-spanning avalanches resulting  in a global "failure" of the pillar; one such system size avalanche is  marked by an arrow on Fig.\ref{fig:mildtosup}(a). Note that  the corresponding mechanical response is brittle-like, with only few inelastic events preceding the  major collapse. This is similar to  the previously studied behavior of compressed nanoparticles in the diameter range 200 to 800 nm.  They show an extreme case of brittleness with a unique system-spanning dislocation avalanche preceded by a purely elastic response \cite{mordehai2011,sharma2018}. However, a quantitative comparison of our BCC pillars (Mo),  fabricated from a bulk sample containing pre-existing dislocations, and such FCC (Au and Ni) nanoparticles, fabricated from solid-state dewetting and initially dislocation-free, has to be done with caution. 

All these   observations argue for a rich spectrum of mechanical behaviors depending on the  crystal  size $L$: from a smooth, ductile-like response associated with non-scale-free (mild) fluctuations in "large"  crystals,  to a purely brittle behavior characterized by a single dragon-king event in "small" crystals. In this context, the critical scale-free dynamics, observed in our Mo pillars for $L$=1000 nm, appears as an intermediate regime   underlying  a brittle-to-ductile (BD) transition which can be then  associated with  particular  range of system sizes. 
 
Interestingly, a BD transition is usually discussed in the context of fracture, with dislocation nucleation at crack tip indicating the emergence of ductility \cite{rice1974, gumbsch1998}. The results presented above argue for a BD transition "without a crack", i.e. in terms of collective behavior of dislocations characterized by a particular scale free structure of plastic fluctuations. In this case, rather paradoxically, the nucleation of individual dislocations is associated with a brittle regime \cite{mordehai2018}. 

\subsection{The concept of wildness}

In the setting discussed  above, the evolution from super- to sub-criticality over a  range of system sizes occurs in a BCC material. 
In  fact, a similar dependence of the  plastic fluctuations on system size $L$ has been also observed in FCC materials. As already noted, initially dislocation-free FCC nanoparticles do exhibit brittleness, even though a supercritical regime has not (yet) been observed  for FIB-fabricated FCC pillars. The latter may be related to the difficulty of  fabricating  dislocation-free crystals due to the  easiness for    Ga$^+$ ions to penetrate  light FCC crystals, which generates FIB-induced defects (small dislocation loops) near the surface  \cite{wirth2007,kiener2007}. 

\begin{figure}[h]
 \includegraphics[width=0.8\linewidth]{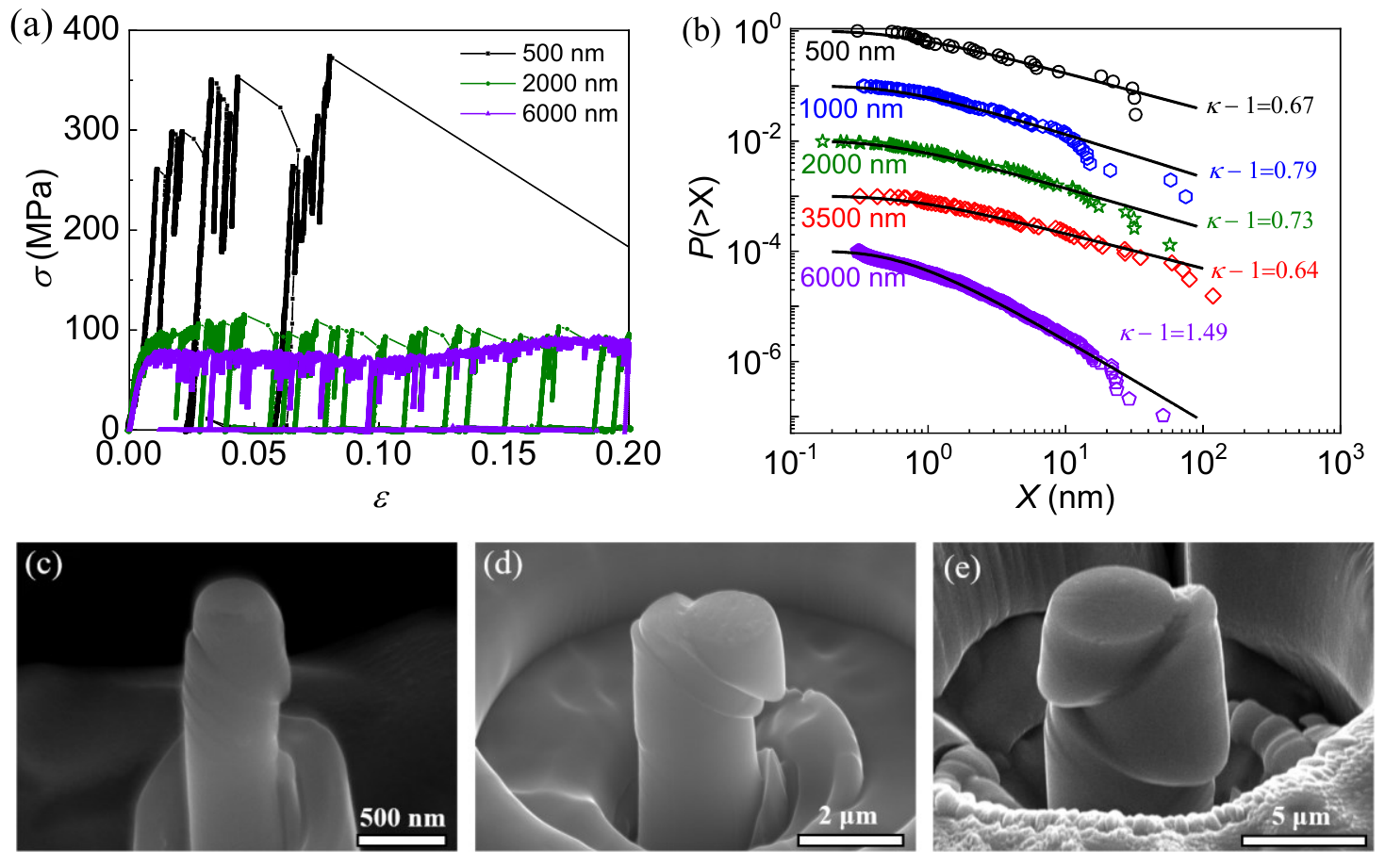}
 \caption{From mild to wild fluctuations in Al micropillars. (a) stress-strain curves (shear). (b) Cumulative distributions of plastic displacements $X$ detected over the entire loading. The solid lines represent the fit of the data with eq. (2), and the corresponding lower cut-off values $X_0$ are, from top to bottom: 0.64, 0.73, 0.61 and 1.02 nm. (c) to (e): SEM images of (c) 500 nm, (d) 2000 nm and (e) 6000 nm micropillars after compression. Multislip is only observed for the largest micropillars (6 µm). }
 \label{fig:Alwildness}
 \end{figure}

However, a recent study \cite{lee2016} showed that such  FIB-induced dislocation loops in Al micropillars can be eliminated by thermal annealing, which makes the stress-strain response more brittle-like. Unfortunately, the associated statistics of strain burst sizes has not been analysed. While our observations on BCC crystals argue for a sharp transition from a critical dynamics at $L\sim$ µm to a more ductile behavior at larger system sizes \cite{brinckmann2008,zhang2017}, similar transition  seems to be more gradual for FCC  materials. Figure \ref{fig:Alwildness} illustrates this effect for pure Al. While the size effect on strength is still apparent (Fig. \ref{fig:Alwildness}(a)), slip burst distributions are characterized by a power law tail over a wider $L$-range ($\sim$ 500 to 3500 nm) than for Mo, with $\kappa\approx$1.6-1.7, closer to the mean-field value 1.5 (Fig. \ref{fig:Alwildness}(b)). At larger system sizes, say $L$=6 µm,   a larger exponent emerges, in association with a smoother mechanical response. These observations may be viewed as another argument supporting  the idea of  non-universal character of the exponent $\kappa$.

One can show that  all the distributions  shown in Fig.\ref{fig:Alwildness}(b) are well described by the generic formula
\begin{equation}
    P(X)=\frac{X_0^{\kappa-1}}{\Gamma(\kappa-1)X^{\kappa}}e^{-X_0/X}
\end{equation}
\begin{equation}
    P(>X)=1-\frac{\Gamma(\kappa-1,\frac{X_0}{X})}{\Gamma(\kappa-1)},
\end{equation}
where $\Gamma(a,x)$ is the incomplete gamma function, and the exponential term represents a \textit{lower} cut-off to power law statistics occurring around $X=X_0$. These expressions characterize the presence of wild (scale free, power-law distributed) fluctuations at large scales $X\gg X_0$, coexisting with mild fluctuations associated with a characteristic size $X_0$ at small scales. Eqs. (1-2) received a theoretical support from a simple mean-field stochastic model of dislocation dynamics \cite{weiss2015}, which we will discuss in some detail later in the paper (see section 5.2). 

Note that for pillar sizes $L\geq1$ µm, see  Fig.\ref{fig:Alwildness}(b), the lower cut-off $X_0$ is not an experimentally imposed threshold. In particular,  it increases for the largest Al pillars. This will become even more apparent  when we consider alloys (see section 3). However, it is clear that  large range of mild fluctuations, with sizes $X<X_0$,  remain  undetected by our  methodology \cite{zhang2017}. This is particularly the case for $L$=500 nm Al pillars and $L$=1 µm Mo pillars, where the  lower cut-off is hardly discernible on our data. 

In addition, we did not find any statistical evidence  of an \textit{upper} cut-off limiting  the power law range in our Al data, independently of  $L$. In particular, for the 1500 nm Mo pillars (Fig. \ref{fig:mildtosup}(b)), eq. (2) with a large $\kappa$ value fits well the data. In that case, however, the generic form given above can be  hardly differentiated from a log-normal distribution. In any case, all this  means  that the observed  sub-critical regime is strongly dominated by mild fluctuations.

The relative contribution of mild vs wild fluctuations to the total plastic strain can be reasonably estimated. We call $X_{min}$, the value, close to $X_0$, above which fluctuations can be considered as being power-law distributed, i.e. wild, and write $K=e^{-X_0/X_{min}}$. Then,  the probabilistic weight of the wild part of the distribution (1), measuring the amount of plastic strain accommodated through scale-free avalanches, which we call \textit{wildness}, and denote by $W$, can be written as (see also section 5.2) \cite{weiss2015}:
\begin{equation}
    W(\kappa,K)=1-\frac{\Gamma(\kappa-1,\ln(1/K))}{\Gamma(\kappa-1)}
\end{equation}
The most striking point here is the link between the wildness $W$ and the power law exponent $\kappa$, with large $W$ being associated with smaller values of exponents. Experimentally, $\kappa$ as well as the lower bound $X_{min}$ can be estimated using the  maximum-likelihood methodology \cite{clauset2009,zhang2017}, and $X_0$ found from a best-fit of eq. (2). The plastic strain dissipated through scale-free avalanches can be independently  calculated by accumulating all the strain bursts above $X_{min}$, then comparing the associated strain  to the total plastic strain in order to estimate $W$ \cite{zhang2017}. We will show later in section 3 that eq. (3) is consistent with experimental data for  a surprisingly broad range  of materials, including pure metals and alloys. 

Here we only mention  that FCC materials are characterized by an extended BD transition taking place with increasing  system size. A new finding is that this transition is associated with a progressive evolution of wildness and exponent $\kappa$. We recall that in such bulk materials macroscopic plasticity is  mostly smooth ("ductile") with only  scarce bursts   detected by AE (see section 2.1). At the macroscopic  scales a proxy of wildness can be estimated from $W_{AE}$, by summing the energies of the detected  bursts and comparing with the total emitted acoustic energy (both continuous and discrete, see section 2.1) \cite{weiss2015,weiss2019}. In  bulk FCC materials, $W_{AE}$ rapidly decreases to a few \% or less as strain hardening takes place and   dislocation substructure form \cite{weiss2015,LHote2019}. This is  particularly visible under cyclic loading where dislocation avalanches become exceptionally rare  already after few cycles \cite{weiss2019b} (see section 4). 

Despite the achieved understanding, the issue remains with BCC materials   exhibiting a sharper BD transition than FCC materials. The origin of this difference will be discussed in the next section.

\subsection{External vs internal length scales}

The results presented above show clear evidence of the strong effect of the  system  size  on plastic fluctuations. The quantitative role of such an external scale  can be studied if we  present it as a ratio of two length scales, $R=L/l$, where $l$ represents some judiciously chosen  internal length scale.  The resulting nondimensional parameter  will then control the  collective dislocation dynamics \cite{zhang2017}. 

In case of the Mo pillar experiments illustrated in Fig.\ref{fig:mildtosup}, the initial dislocation density  before loading was measured giving $\rho=1.6\times10^{12}$ m$^{-2}$. This  implies that the  mean dislocation spacing is  $l_d$= $1/\sqrt{\rho} \approx$ 790 nm.  We can interpret the length $l_d$, which is  naturally present in all pure materials, as an internal scale. 

With this interpretation at hand, we note that  supercriticality is observed for $L<l_d$ meaning $R<1$ and subcriticality for $L>l_d$ or $R>1$. This is not incidental, and illustrates the role of the  mutual interactions of dislocations (short-range and long-range) in setting the internal dynamics.

Indeed, for $R$<1, preexisting (and mostly locked) dislocations, serving as potential obstacles in ensuing dislocation dynamics,  are almost absent.  Therefore the work of the loading device  is dissipated through a  collective nucleation and correlated motion of unlocked  dislocations. In particular, such lack of inhibition  leads to the emergence of super-critical instabilities. 

Instead, for $R>>$1, moving dislocations  necessarily encounter considerable number of different obstacles, in particular,  forest dislocations.   In this context, cross-slip as well as short-range interactions are promoted, frustrating the development of system size  correlated dislocation avalanches. Hence, mild fluctuations and a ductile-like behavior ensue. In BCC materials large lattice friction will likely play a role as well in inhibiting large avalanches (see more on this below). 

Over an intermediate  range,  $R \sim 1$, i.e. for $L \sim l_d$, we observe that while the pillars are not initially dislocation-free, short and long-ranged (elastic) interactions are nicely balanced, which apparently creates the right  environment for self organization to a scale-free, critical avalanche dynamics. The deformation morphologies shown on Fig. \ref{fig:mildtosup}(c) to (e) support this interpretation: at small pillar sizes, plastic deformation is highly anisotropic with a clear domination of a single slip plane, while at large pillar sizes, the plastic flow is   more diffuse and isotropic as the result of the dominating multislip deformation.

A natural way to define the internal length scale $l$ in the general case is to write \cite{weiss2006,kubin2013,zhang2017}:
\begin{equation}
    l=Gb/\tau_{pin},
\end{equation}
where $G$ is the shear modulus and $b$ the Burgers's vector. The new parameter $\tau_{pin}$ is the effective pinning strength of  the existing obstacles inhibiting dislocation motion. We can write 
\begin{equation}
    \tau_{pin}=\tau_l+\tau_f+(\tau_{s}+\tau_{p}),
\end{equation}
where $\tau_l$ is the lattice friction, $\tau_f=\alpha_f Gb\sqrt{\rho_f}$ is the pinning strength of forest dislocations (with  constant $\alpha_f<1$).   The two other  terms, $\tau_{s}$ and $\tau_{p}$,corresponding  to extrinsic disorder  and describing the effects of solutes and  precipitates, respectively,  will be discussed in section 3 devoted to  alloys. Given this definition, the internal scale $l$ corresponds to the distance  at which the dislocation/dislocation elastic interaction stress (scaling as $Gb/l$) becomes comparable to the dislocation/obstacle interaction stress $\tau_{pin}$.

\begin{figure}[h]
 \includegraphics[width=0.8\linewidth]{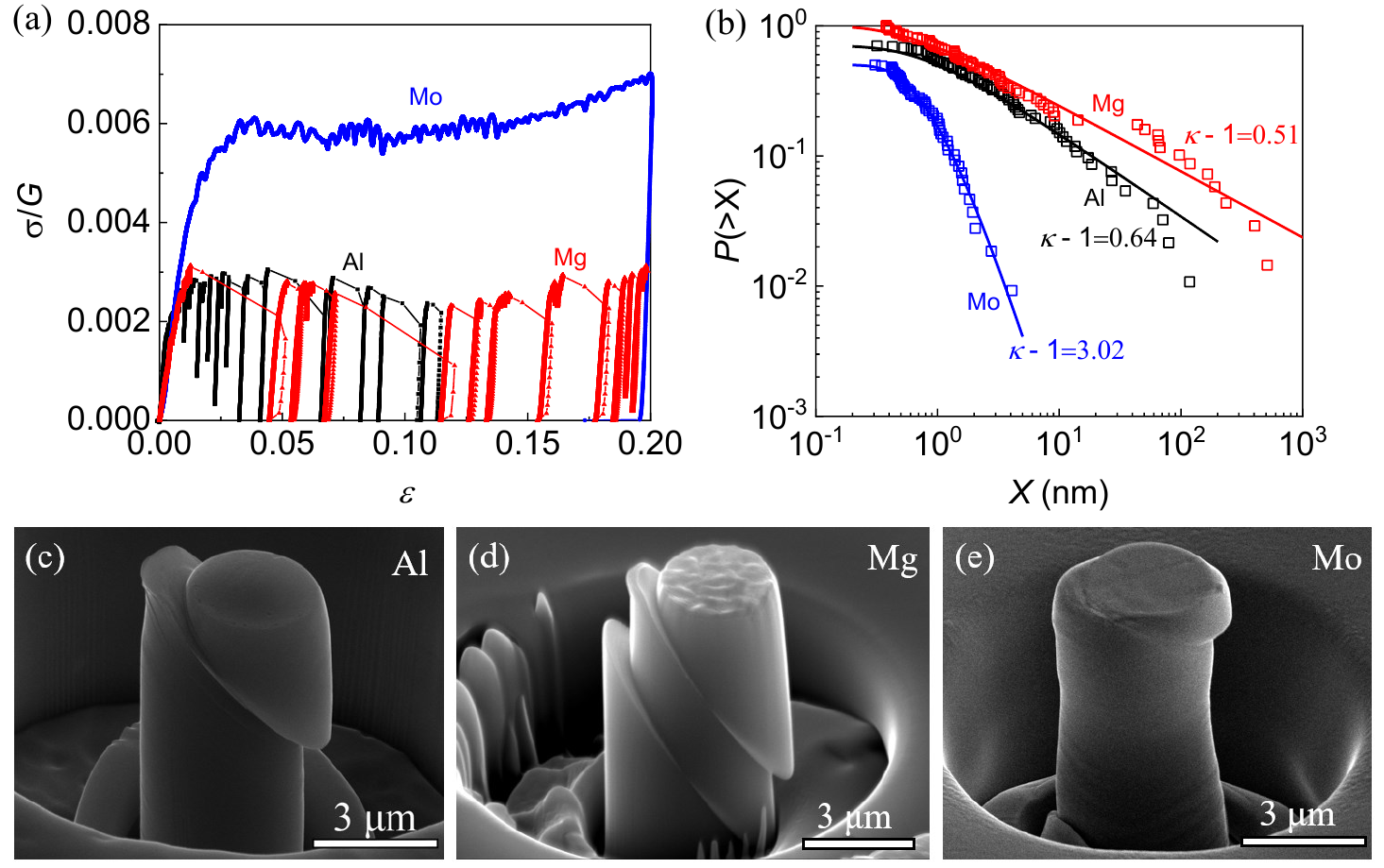}
 \caption{The effect of lattice structure on plastic fluctuations. (a) A comparison of stress-strain curves for 3500 nm pillars of three different materials, Mg (HCP), Al (FCC) and Mo (BCC). The stresses have been normalized by $G$, the shear modulus of each material. (b) Corresponding cumulative distributions of plastic displacements $X$ over the entire loading. The solid lines represent the fit of the data with eq. (2), and the corresponding lower cut-off values $X_0$ are 0.50 nm (Mg), 1.02 nm (Al) and 2.07 nm (Mo). (c) to (e): SEM images of (c) Al, (d) Mg and (e) Mo micropillars after compression. Single slip is observed for Al and Mg, while isotropic deformation is observed for Mo. }
 \label{fig:lattice}
 \end{figure}

Experimentally, $\tau_{pin}$ can be in principle obtained from the yield strength under tension of bulk samples \cite{zhang2017}. However, caution is necessary when we deal  with small scales. Consider e.g. BCC materials below an athermal transition temperature $T_a$.  Then the lattice friction experienced by screw dislocations is large  as the result of their non-planar core structure. This  implies the necessity  to overcome large barriers  through thermal activation.  Our tests on Mo, performed at room temperature, belong to this range, as $T_a\approx$ 465 K in this material \cite{kubin2013}. In FCC materials,  where lattice friction is small, independently of  the temperature, one can  expect a smaller value of $\tau_l$ than in the case of BCC. This would lead to   smaller value of the internal scale $l$  and will reduce the $L$-range over which large fluctuations are expected. To confirm this conjecture we compare in Fig. \ref{fig:lattice}  the data for  Mo (BCC), Al (FCC) and Mg (HCP) pillars of the same size (3500 nm).  In particular, the Mo pillars display a much smoother mechanical response and a larger exponent $\kappa \approx 4.0$. Given this scenario, one would expect the difference between FCC and BCC to disappear for $T>T_a$, when the role of thermal activation in the dynamics of  screw dislocation disappears. This assertion is supported by the results reported in  Abad et al.\cite{abad2016}, where  stronger plastic intermittency in BCC materials at higher temperatures was observed. At these temperatures  lattice friction apparently  no longer inhibits large scale self organization and the nature of the internal scale $l$  changes.

Another factor biasing the  competition between short-range and long-ranges interactions is the damping of dislocation motion: if strong enough, it  can inhibit fast  dislocation avalanches \cite{cui2020}. Consistently with this observation, plastic intermittency is suppressed in micropillars deformed at  strain-rates ($\geq 1$ s$^{-1}$) that are larger than the internal relaxation rate limited by lattice friction \cite{sparks2019}. The damping related effects should not, however,  depend on system size $L$,  which excludes its role  in the critical to super-critical transition observed in BCC micropillars even at low $T$ (Fig. \ref{fig:mildtosup}). Here it is appropriate to refer   to recent DDD simulations which suggest that at low $T<T_a$ and relatively large system sizes, strain fluctuations are controlled by the slow (thermally activated) screw dislocation motion, hence avalanches can be damped \cite{cui2020}. Instead, at sub-µm sizes, the  external stress imposed on samples enhance the athermal mobility of screw dislocations to the level of edge ones,   making irrelevant the  thermally activated  motion. In particular, the DDD simulations show a reduced sensitivity of strength to temperature as the size of BCC micropillars diminishes \cite{cui2016}. This implies that the damping   mentioned above is  suppressed, which allows the initiation and propagation of strain bursts \cite{cui2020}. To summarise,  switching from  long- to  short-ranged interactions  and   variable  role of damping may be the factors explaining the sharper BD transition observed in BCC compared to FCC.       
 
Questions, of course, remain regarding the interpretation of the internal scale $l$ as defined by eq. (4). Thus, in BCC materials, dislocation patterning and cell formation are not observed below a certain value of applied strain  which strongly increases with decreasing temperature. This is related  to  temperature dependence of lattice friction and the limited ability of screw dislocations to cross-slip at low temperatures \cite{keh1963,kubin2013}. In this case, we can define  $l_d=1/\sqrt{\rho_f}$, which means that we link the internal scale with the dislocation mean free path. However, the relation between $l$ and $l_d$ over a large range of conditions is  not  simple.  On the one hand, increasing lattice friction enlarges  $\tau_{pin}$, hence lowers $l$.  On the other hand, the constant $\alpha_f$ in Taylor's relation, $\tau_f=\alpha_f Gb\sqrt{\rho_f}$, is smaller than 1. 

In FCC materials, lattice friction plays a minor role, and  the link between $l_d$ and a dislocation mean free path is less direct (e.g. \cite{zaiser2014}). Indeed, dislocation patterning emerges naturally in these materials and the corresponding characteristic scale $l_p$ (e.g. cell size) appears as a natural mean free path. The "similitude  principle" then states \cite{sauzay2011,kubin2013}
\begin{equation}
    \frac{\tau_{pin}}{G}=k\frac{b}{l_p},
\end{equation}
which, combined with (1), relates $l_p$ to our  internal length scale
\begin{equation}
    l_p=k\frac{Gb}{\tau_{pin}}=kl,
\end{equation}
where $k$ is a dimensionless constant;
a comparison of experimental data for different materials suggests  that    $k\simeq$7.5 \cite{kubin2013}.   

Similar reasoning can also help to rationalize the effect  of crystal symmetry on plastic intermittency and the associated   dislocation morphologies. We have already discussed above the systematic differences between FCC and BCC crystals. Figure \ref{fig:lattice} shows  the results for HCP material. We deal here with 3500 nm pillars of Mg. The observed mechanical response is extremely jerky, represented by a  succession  of plastic avalanches, which are power-law distributed in size with $\kappa\simeq$1.5. These observations for micropillars are  entirely consistent with the recorded  behavior of HCP bulk materials, where plastic jerkiness has been noticed long ago. It was seen  directly on stress strain   curves obtained for  thin rods \cite{becker1932,tinder1973}, and was later quantified based on  AE measurements for general samples \cite{miguel2001,richeton2006,weiss2015,weiss2019}. All these studies revealed a high degree of wildness $W_{AE}\simeq$1 independently of which  HCP single crystals were studied (ice, Zn, Cd). The  associated   exponent for the distribution of AE energies was found to be very close to the mean field value 1.5

These observations suggest an absence of the  size induced BD transition for HCP materials. Apparently, they  remain in a critical state starting from the  system size of a few µm and all the way to bulk scales. To rationalize these observations  within our framework, we recall that  the mentioned HCP materials  are characterized by a very strong plastic anisotropy: gliding in such materials  takes place  preferentially along the basal planes whatever the system size.  Another characteristic feature is a low Peierls stress and the associated lack of   lattice friction. In other words, our parameter  $\tau_l$ in such materials is anomalously small \cite{kubin2013,weiss2019}. Plastic anisotropy implies an absence of forest hardening, which means that  $\tau_f$ is also negligible.  Altogether this means that $\tau_{pin}$ is sufficiently small for  $l$ to be  comparable to $L$. Then    long-ranged elastic interactions fully control the collective dynamics and nothing prevents self organization of   dislocations  towards  a critical state along the whole range of system sizes. The above  arguments can explain the absence of a ductile/subcritical regime.  The possibility of a super-critical regime in these materials is also  very small because it  requires  initially dislocation-free environment   which can  be achieved only if  the system size is  decreased well below few µm.

The plastic anisotropy of Mg pillars is   apparent in Fig. \ref{fig:lattice}(d) showing a single slip morphology. In fact, a strong correlation between plastic anisotropy and criticality should be kept in mind. As we have seen  whatever the material and its crystal symmetry, either  critical or supercritical regimes strongly correlate with slip anisotropy (see Figs. \ref{fig:mildtosup}(c and d), \ref{fig:Alwildness}(c and d), \ref{fig:lattice}(c and d)). Instead,  ductile behavior can be  associated with  homogeneous deformation resulting from  isotropization due to  multislip (e.g. Fig. \ref{fig:mildtosup}(e) and \ref{fig:lattice}(e)). In other words, there is a  strong link between the dominance of short-range interactions, multislip and ductility (or subcriticality).   In contradistinction,  initial  purity,  dominant long-ranged interactions and  extreme slip anisotropy, correlate with brittleness  (or supercriticality). In these systems, however, supercriticality easily turns into a scale free behavior over the whole range of observable in the presence of a non-vanishing initial dislocation density.

In the next section we show  how the above arguments transform  in the context of  plastic flows in alloys (see below) \cite{zhang2017}.     

\section{Alloys}

The introduction of impurities has been used for millennia in classical metallurgy to harden bulk materials (e.g. \cite{verhoeven1998}).  Following this logic, the "smaller is stronger" size effect occurring in pure metals at µm and  sub-µm scales, might also appear beneficial. However, as we detailed above, such effect  at small scale is accompanied by a detrimental "smaller is wilder" effect. The latter is culminating in brittleness at nanoscale (see above), which may compromise  the forming processes and endanger the load-carrying capacity in various engineering/industrial processes taking place in such samples \cite{gao2014,hu2016}. A key challenge is therefore to develop new metallurgical procedures to mitigate the associated  plastic instabilities  and reduce/suppress their detrimental effects, while preserving all the  advantages related  to high  strength.

Studies revealed a much weaker size effect on strength in alloys compared with pure materials. This was  interpreted as the fingerprint of a tailored internal length scale, which also suggested the possibility of the  reduction of plastic intermittency \cite{gu2013,girault2010,pan2019}. The framework proposed in section 2.4, and particularly our eqs. (4) and (5), can be used to  rationalize these observations.

Indeed,  the observations suggest  an increase of the mean-field pinning strength $\tau_{pin}$ as the addition of solutes ($\tau_{s}$) or precipitates ($\tau_{p}$) should reduce the scale $l$. Since the latter controls   the transition from a long-ranged to a short-range controlled dynamics, alloying  expands the range of system sizes $L$ over which plastic fluctuations remain mild. 

\begin{figure}[h]
 \includegraphics[width=0.8\linewidth]{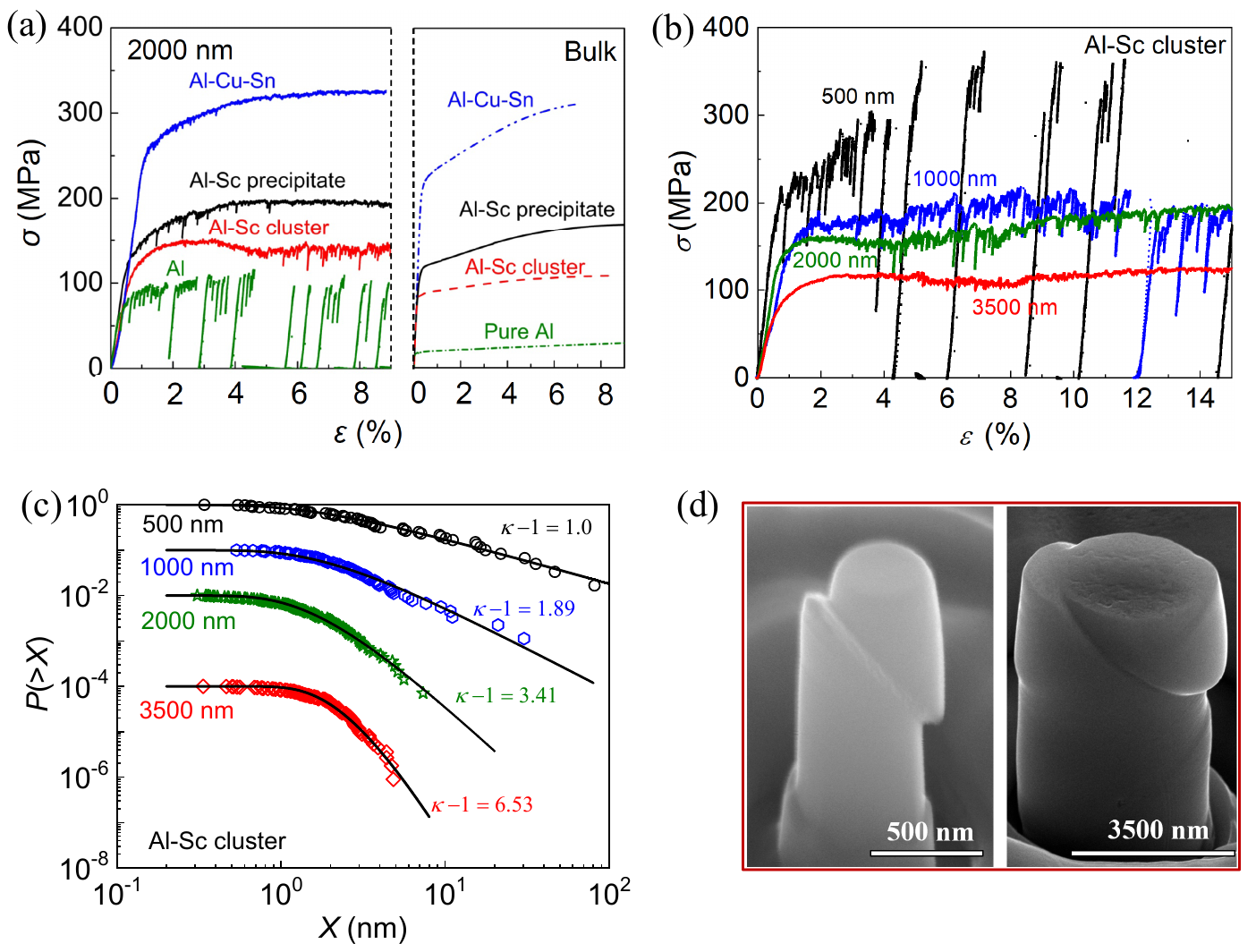}
 \caption{Taming plastic fluctuations from alloying: the case of Al-Sc alloys. (a) Stress-strain curves for pure Al and Al alloys. (b) Stress-strain curves for Al-Sc cluster pillars of different diameters. (c) Corresponding cumulative distributions of plastic displacements $X$ over the entire loading. The solid lines represent the fit of the data with eq. (2), and the corresponding lower cut-off values $X_0$ are 1.91 nm (500 nm pillar), 3.16 nm (1 µm pillar), 4.11 nm (2 µm pillar), and 11.2 (3.5 µm pillar). (d) SEM images of a 500 nm pillar (left) and a 3500 nm pillar (right). Single slip is observed for the small pillar, and multislip for the large one. Adapted from \cite{zhang2017}.}
 \label{fig:Al-Sc}
 \end{figure}

Our own experiments on Al alloys micropillars also confirmed the correctness  of this interpretation \cite{zhang2017}. Micropillars  containing either clusters of Sc solutes (so-called Al-Sc cluster alloy), Al$_3$Sc precipitates (Al-Sc precipitate alloy), and  $\theta'-$Al$_2$Cu plate-like precipitates (Al-Cu-Sn alloy) were tested. They are characterized by  an increasing pinning strength $\tau_{pin}$ that can be estimated from tensile tests at bulk scale. Figure \ref{fig:Al-Sc}(a) clearly shows that such alloying can strongly reduce wild, unwelcome plastic fluctuations, while preserving a strengthening effect. At the level of an individual  alloy, the "smaller is wilder" effect is still present (Fig. \ref{fig:Al-Sc}(b)). Therefore, even if it has been shown  that alloying can weaken or even suppress the size effect on strength \cite{girault2010,gu2013}, an effect on plastic fluctuations remains.  However, in alloys this effect is  definitively shifted towards smaller system sizes comparing to pure materials. 

The distributions of displacement burst sizes $X$ in the studied  alloys  were shown to be of mixed character, very well described by the generic expressions (1-2). The associated exponent $\kappa$ and  the characteristic size $X_0$ increase with increasing sample size $L$ and pinning strength $\tau_{pin}$ (Fig. \ref{fig:Al-Sc}(c)). We did not detect any signs  of super-criticality over the analysed $L$-range which  is not surprising, given that  the associated  transition towards pure brittleness is  not observed in our pure Al samples either, down to $L$= 500 nm. 

\begin{figure}[h]
 \includegraphics[width=0.8\linewidth]{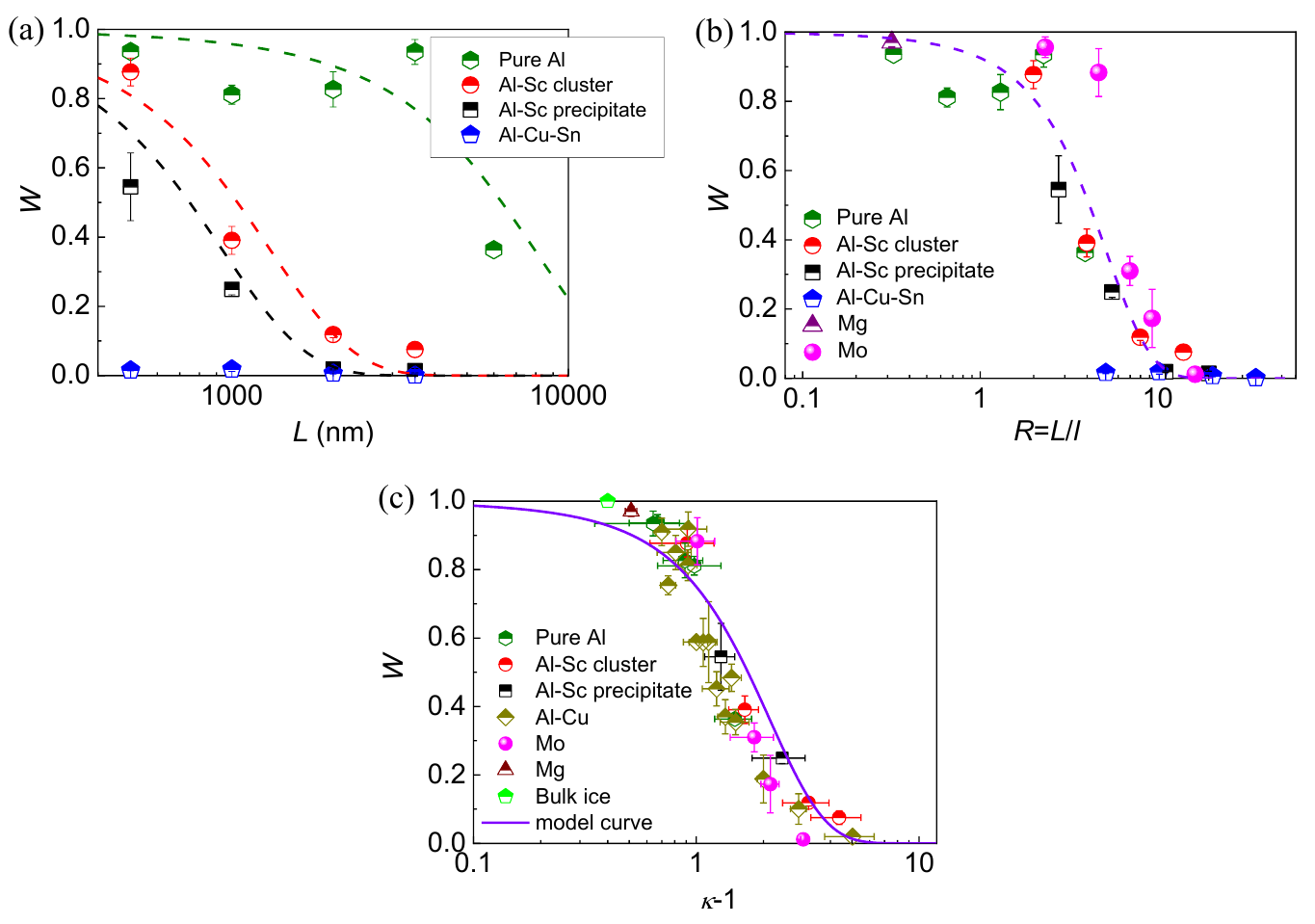}
 \caption{The universal character of the wild-to-mild transition. (a) Wildness $W$ as a function of pillar diameter $L$ for pure Al and various Al-Alloys with increasing pinning strength. (b) Wildness $W$ as a function of the dimensionless ratio $R=L/l$ for different pure materials and alloys, where $l$ was computed from eq. (4) using $\tau_{pin}$ values estimated from tensile tests at bulk scales, except for Mg for which the lattice resistance of basal slip was used, i.e. $\tau_{pin}=\tau_l=$0.5 MPa. (c) The universal relationship between wildness $W$ and the exponent $\kappa$.}
 \label{fig:universal curve}
 \end{figure}

In section 2, we argued that the ratio of length scales $R=L/l$ is the key controlling factor of the wild-to-mild transition.  We also concluded that there exists  a universal (material-independent) relationship between wildness $W$ and the corresponding exponent $\kappa$ (eq. (3)). Figure \ref{fig:universal curve} shows the remarkable precision  of these predictions.

When plotted as a function of the system size itself, the values of wildness parameter  get shifted towards smaller $L$ when we increase  the pinning strength of extrinsic disorder (Fig. \ref{fig:universal curve}(a)). However, a normalization of $L$ by the internal scale $l=Gb/\tau_{pin}$ collapses the data for pure Al and for the  different alloys of Al on the same curve.  It shows that a wild-to-mild transition indeed takes place  around $R\sim$ 1  (Fig. \ref{fig:universal curve}(b)). This prediction was also checked  for the other materials tested, in particular, for  Mo (BCC metal), which confirmed its generality. Fig. \ref{fig:universal curve}(c), compiling $W$ and $\kappa$ values obtained for sample sizes ranging from sub-µm to bulk scales (in case of ice), various pure materials with different crystal symmetries, as well as different FCC (Al) alloys, illustrates the robustness of eq. (3). The obtained agreement with observations indicates a sound basis of the underlying theoretical ideas.   While the  individual scaling exponents turn out to be  nonuniversal \cite{weiss2015,zhang2020} the ensuing dependence of exponents on wildness appear to be universal. The theoretical ideas  behind the proposed  framework which justify such a generalized notion of universality will be discussed in section 5. 

Here, we also mention that that the correlation between the anisotropy of the deformation morphology and the wildness,  observed in pure materials, is also recovered  in alloys (Fig. \ref{fig:Al-Sc}(d)). In particular, the extrinsic factors which contribute to isotropy and promote multi-slip, also  decreases the wildness\cite{zhang2017}. These results suggest that wild plastic intermittency, implying unwelcome large fluctuations,  can   be systematically subdued from a tailored set of homogenizing defects, at least in FCC materials. The possibility of  extending  these wildness suppression techniques to other crystalline structures, and, in  particular to HCP materials displaying plastic jerkiness even at macroscales \cite{becker1932,tinder1973,weiss2019}, remains to be explored.

In the studied FCC alloys, $\tau_{pin}$, and therefore the internal scale $l$, are largely controlled by the extrinsic disorder terms $\tau_s$ and $\tau_p$. These parameters  depend on both, the pinning strength of individual obstacles and their average spacing $\lambda$. In the studied samples the  spacing $\lambda$ was always smaller than $l$, sometimes (Al-Sc cluster alloy) by more than an order of magnitude. Therefore,   $\lambda$ is not an appropriate normalization scale to reveal the  underlying wild-to-mild universality.  In all cases, $\lambda$ was also much smaller than $L$, justifying our approximate formula   for $\tau_{pin}$. One  can wonder what would  happen when $\lambda  \sim L$, i.e. when dislocations can  potentially cross over the system without meeting an obstacle. Will this  eliminate the beneficial effect of alloying and open the way towards supercriticality and brittleness ? 

While these questions remain essentially open.  we can refer to our  recent work on Al-Cu alloys with much larger $\theta'-$Al$_2$Cu plate-like precipitates, i.e. of diameter $d_p$ commensurate with the  size of the micropillars \cite{zhang2020b}. In that case, our simplified  picture breaks down. In particular, comparing to  alloys with much smaller obstacle sizes but a similar $\tau_{pin}$ (deduced from tensile test at bulk scales), the ensuing plastic flows are much  wilder. A sharp decrease of wildness was  observed over an intermediate range of system size, when $d_p \simeq L$ and the precipitates cut the entire micropillar. Despite the implied   modification  of the pinning picture, the statistic of plastic fluctuations in these alloys remained  consistent with the proposed $W$ vs $\kappa$ relationship (eq. (3)), which should be viewed as another manifestation of its  universal character. 

\section{Wildness  vs   strength}

In the discussion above, we assumed that in pure materials with low lattice friction, the parameter $\tau_{pin}$ is mainly controlled by forest dislocations. More specifically, we argued that the  term $\tau_f \sim Gb / \sqrt{\rho_f}$ must be  tightly linked to Taylor's forest hardening. This idea is supported by the extreme wildness of pure divalent HCP metals (Mg, Zn, Cd) and  ice, even at macroscopic scales \cite{becker1932,tinder1973,miguel2001,richeton2006,weiss2019}.  It  is illustrated in Fig. \ref{fig:universal curve}(c),  where we see that  $\kappa \simeq 1.5$ and $W \simeq$ 1. We recall that in such materials, plastic deformation is strongly anisotropic, with preferential glide along the basal plane and a correspondingly very small Peierls stress $\tau_{l}$ \cite{kubin2013,weiss2019,bacon2002}. The strong plastic anisotropy implies an absence of forest hardening, i.e. a possibility to completely neglect  $\tau_f$. In fact, both effects conspire to ensure a large value of $l$ and therefore  a wild behavior over an extended range of system sizes. 

Note, however,  that other HCP metals, such as Ti and Zr, preferentially glide along prismatic planes. They also exhibit  a considerable temperature-dependent lattice friction for screw dislocations below an athermal transition temperature \cite{kubin2013}. The impact of these and other specific features  of some HCP materials on the jerkiness of plastic flow remains to be explored.

\begin{figure}[h]
 \includegraphics[width=0.9\linewidth]{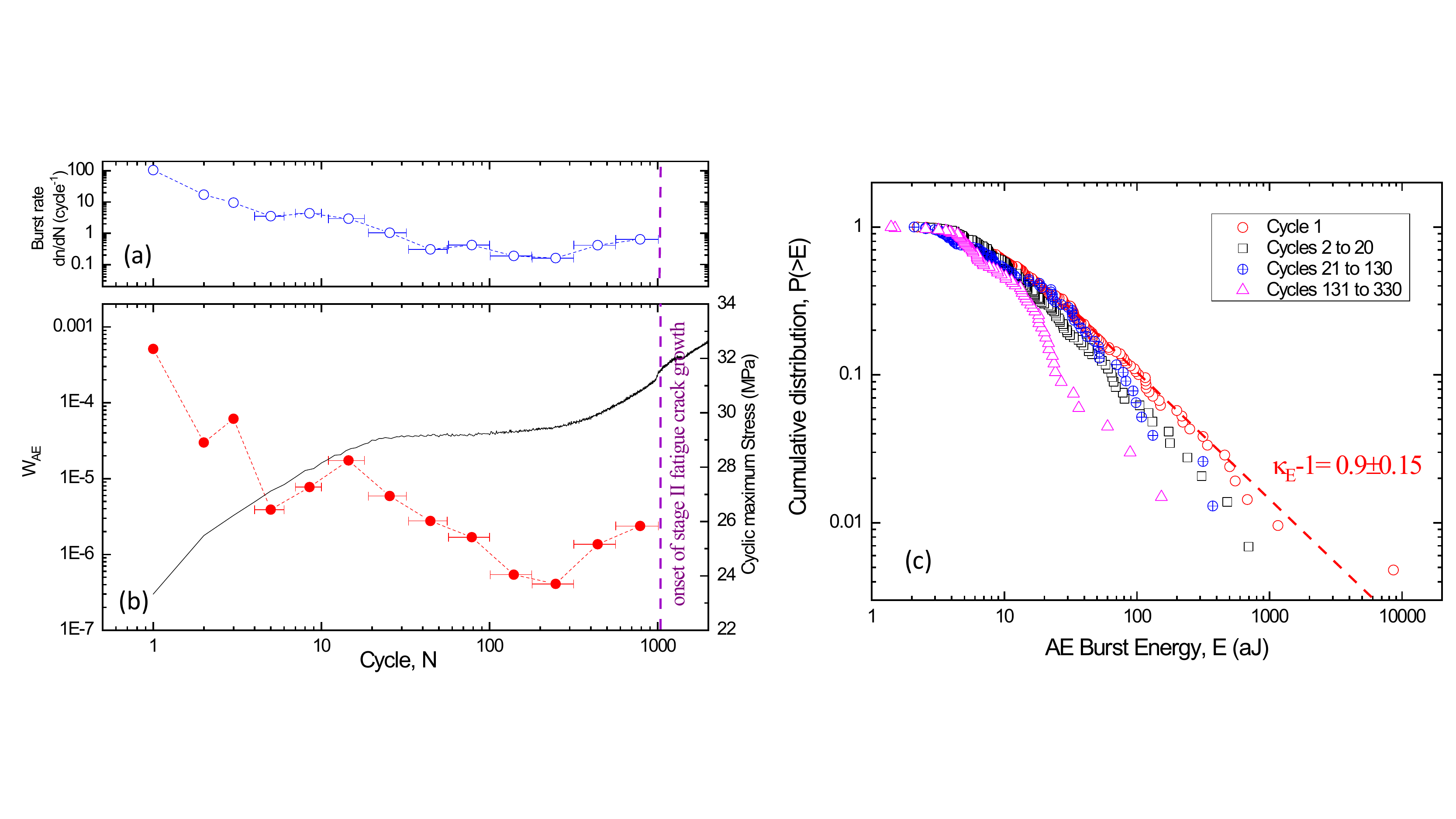}
 \caption{Cyclic strain-controlled tension-compression test ($\epsilon_{min}/\epsilon_{max}=-1; \Delta\epsilon=0.95\%$) on an annealed Al sample. The cyclic stress response (csr-curve) is shown as a black line on panel (b), showing an initial hardening stage over the first $\sim$ 30 cycles. The AE burst rate (number of discrete bursts per cycle) is shown on panel (a), while the wildness proxy $W_{AE}$ is shown on panel (b) with red closed circles. (c) Cumulative probability distributions of AE burst energies at different stages of cyclic deformation and hardening. Adapted from \cite{weiss2019b}.}
 \label{fig:cyclic loading}
 \end{figure}

On the other hand, FCC metals are the paradigmatic  forest-hardening materials.  Here one can expect a strong link  between dislocation patterning, hardening rate, and plastic fluctuations and for pure Al \cite{weiss2019b} and Cu \cite{LHote2019} the implied interrelationships were recently analyzed at bulk scales.  Cyclic loading was imposed and  plastic bursts were tracked  with  wildness estimated from AE. As illustrated in Fig. \ref{fig:cyclic loading}, in Al,  both, the   number of detected AE bursts per cycle and   the wildness measure $W_{AE}$,  are  correlated with   hardening.  More specifically, in these cyclic strain-controlled tension-compression tests, the initial cyclic hardening stage, lasting about 30 cycles, is accompanied by a strong decrease of the burst activity (Fig. \ref{fig:cyclic loading}(a)) and the drop of $W_{AE}$ (Fig. \ref{fig:cyclic loading}(b)). Instead,  the  exponent of the power law tail of AE energy distribution progressively increases (Fig. \ref{fig:cyclic loading}(c)). This is consistent with our theoretical predictions, as strain hardening and the associated  increase of  $\tau_f$ suggest  the decrease of $l$ which makes  the plastic flow milder. The results shown on Fig. \ref{fig:cyclic loading} were obtained for annealed Al samples. Measurements performed on non-annealed samples with a larger initial forest dislocation density revealed a similar trend, however with a much smaller initial (1$^{st}$ cycle ) wildness $W_{AE}$, still in agreement with our picture \cite{weiss2019b}.

 Cyclic stress-controlled tests performed on pure Cu, combined with Electron Back-Scattered Diffraction (EBSD) and Rotational-Electron Channeling Contrast Imaging (R-ECCI) observations, shed additional light on the relationship between dislocation patterning and plastic fluctuations \cite{LHote2019}. These tests involved several cyclic steps with increasing stress amplitudes, which were performed on a same sample. During a single step with fixed stress amplitude, the evolution of plastic fluctuations was fully consistent with what we described above for Al, including strain hardening and progressive formation of a dislocation substructure accompanied with a decrease of wildness. A correlation was observed between the dislocation mean-free path $l_p$, estimated from R-ECCI at the end of each cyclic step, and the value deduced from the continuous AE.  The latter was interpreted as  resulting from the cumulative effect of numerous uncorrelated dislocation motions over sweeping areas $A \sim l_p^2$. 
 
 However, it was noticed that upon increasing the stress amplitude, the previously built  dislocation patterns are  destroyed while new ones are rebuilt. Such major restructuring takes place over just  few loading cycles.  The associated  destabilization occurred, at least partly, through dislocation avalanches that  propagated much further than the scale  $l_p$ generated  at the previous stage. In addition, in all the cyclic experiments mentioned above, rare AE bursts were recorded even during the stage of hardening saturation presumably associated with very stable dislocation substructures. This suggests an ultimately metastable character of these patterns, making them  susceptible to episodic large rearrangements spanning over scales much  larger than $l_p$ \cite{weiss2015,weiss2019b}. 
 
Based on  the results obtained in a continuum model of plasticity \cite{zaiser2007} and the  associated simulations in the framework of discrete dislocation dynamics (DDD)  \cite{csikor2007},  it was proposed   that strain hardening  introduces an \textit{upper} cut-off $s_*$ to the distribution of dislocation avalanche sizes. It would then  write $P(s) \sim s^{-\kappa}f(s/s_*)$, with $f(x)$ a cut-off function rapidly decaying for $x$>1. The prediction was that $s_*$ is inversely proportional to the system size $L$ (finite-size effect) and to the hardening coefficient. 

This conclusion can  justify  the decrease  of the energy released in plastic avalanches as the material strain-hardens. However, it  does not explain the concurrent proliferation  of mild fluctuations associated with a degeneration of the \textit{lower} size tail of  the power law avalanche distribution. In fact,  the  effect of hardening on this lower cut-off avalanche scale ($X_0$ in eq. (1))  remains an open question.
 
 \begin{figure}[h]
 \includegraphics[width=0.5\linewidth]{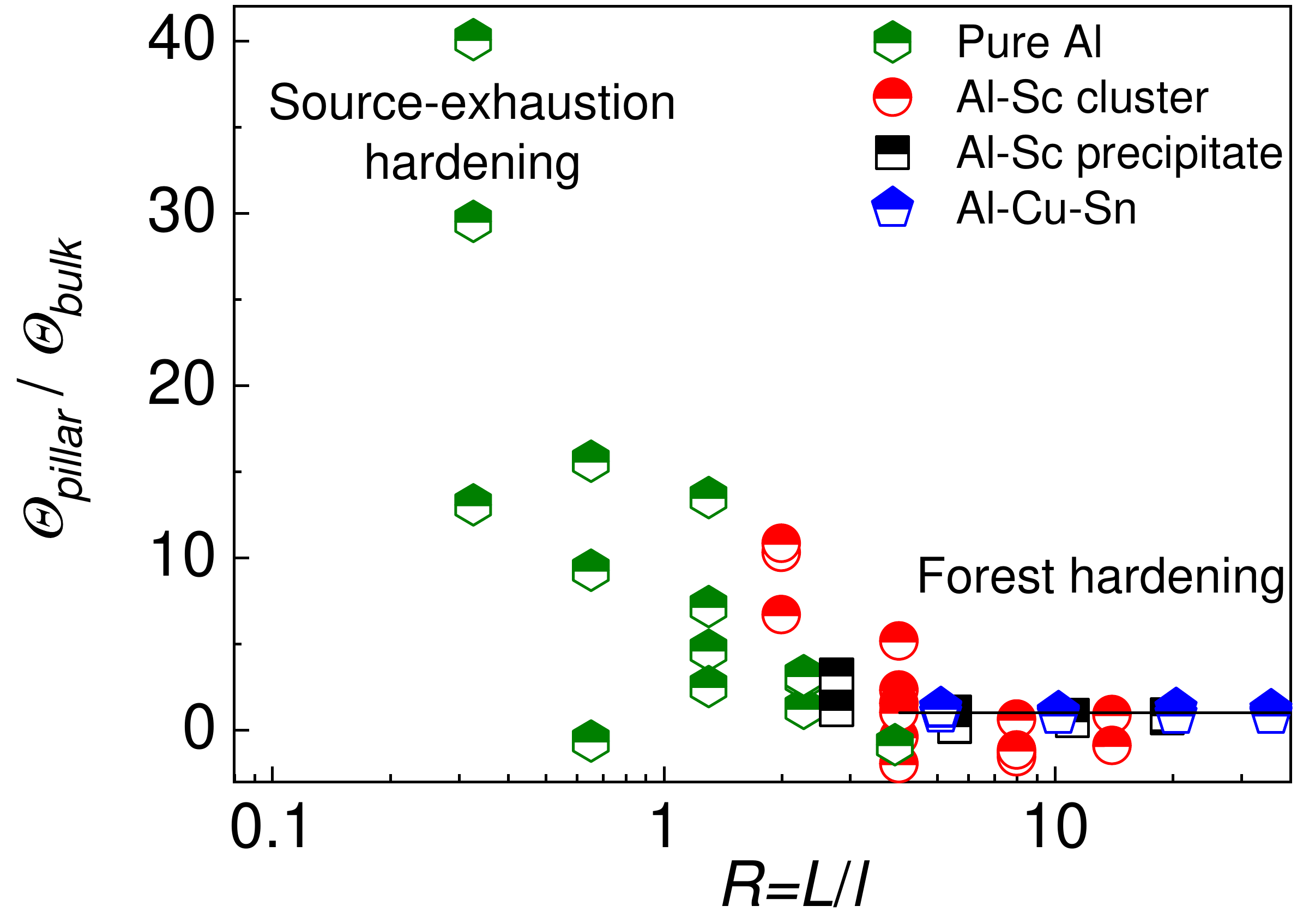}
 \caption{The hardening transition in FCC (Al and Al alloys) micropillars. The normalized strain-hardening rate $\Theta_{pillar}/\Theta_{bulk}$ is shown a function of the dimensionless ratio $R$. A transition from forest hardening to source-exhaustion hardening is observed around $R \simeq$5, whatever the material, in excellent agreement with the mild-to wild transition in Fig. \ref{fig:universal curve}(b). From \cite{zhang2017}.}
 \label{fig:SH}
 \end{figure}
 
 In the considerations presented  above,  strain hardening in FCC bulk materials was analyzed  in the conditions where Taylor's forest hardening was the most  relevant mechanism.  However, upon decreasing the system size below few µm, a breakdown of this  size-independent mechanism can be  envisaged. It can be expected to be replaced by source-dominated mechanisms responsible for the "smaller is stronger" size effect \cite{el2015}. In this case, the usual weak bulk  dislocation sources are almost absent  and much higher stresses are required, in average, to activate the much stronger, surface controlled sources \cite{greer2006}. In such regimes, where isolated breakthrough events dominate,  an increasing scatter of  strength measurements can  be also expected. 
 
 This transition from bulk to surface sources can be  linked to the transition from short-range controlled (allowing forest hardening) to long-range controlled (through distant surfaces) dynamics.  Moreover,  both transitions can be now interpreted in terms of our dimensionless ratio $R=L/l$ \cite{alcala2015}, which was shown to regulate the apparently unrelated  mild-to-wild transition.
 
In this perspective, one can anticipate a relation between the disappearance of forest hardening, the emergence of a size effect on strength, and the new  "smaller is wilder" size effect. The  analysis of compression tests on Al and Al-alloys micropillars allowed us to actually establish such a relation \cite{zhang2017}.

More specifically, we compared the strain hardening rate (SHR) of our pillars, $\Theta_{pillar}$, with those for the same material at bulk scales, $\Theta_{bulk}$. We observed a ratio $\Theta_{pillar}/\Theta_{bulk} \simeq$ 1, i.e. a persistence of Taylor's hardening,  down to $R \simeq$ 5, but much larger and more scattered  values at smaller system sizes (Fig. \ref{fig:SH}). Characteristically, the associated transition took place  concomitantly with the mild-to-wild transition (Fig. \ref{fig:universal curve}). The observed  correspondence  extends also to alloys where  extrinsic disorder shifts the transition from forest to source exhaustion hardening towards smaller system sizes. Since the effect of disorder is the same on the mild-to-wild transition, one can argue for the close relation between the  underlying transition.  The disappearance (or at least the  weakening) of size effect on yield stress in alloys in the µm  system size range \cite{gu2013,girault2010} is also fully consistent with this scenario. 

Finally, we mention that with decreasing  the system size further, another relation  between the  fluctuations and the size effects on strength apparently  emerges. Thus, in Au (FCC) nanoparticles  with $L \simeq$ 400 nm,  a saturation of the size effect on plastic yield was observed  together  with a brittle-like behavior (supercritical in our terms; see section 2.2). These observations were interpreted in terms of  a transition from source-exhaustion/truncation hardening mechanism of plastic flow to a homogeneous  dislocation nucleation mechanism \cite{flanagan2019}. Not surprisingly,  in such  regimes the yield strength was found to approach the theoretical strength of the material.  
 
\section{Modelling}

The discussion above highlighted a rich landscape of plastic behaviors in crystalline materials, with the mechanical responses and the  associated fluctuations depending on crystal symmetry, system size, and disorder (either quenched: solutes, precipitates,.., or emergent: forest dislocations, dislocation patterns,..). To rationalize these observations we consider below two types of modelling approaches: mesoscopic and mean field.    

\subsection{Mesoscopic model}

We have shown above that the dimensionless ratio $R=L/l$ appears as the key controlling parameter, encompassing the external size effect ("smaller is wilder") as well as the disorder  through the internal scale $l \sim 1/\tau_{pin}$ \cite{zhang2020}. Below, we show  how  the effects of system size and disorder can be  analyzed in a single setting using a minimal   model of crystal plasticity. The main idea behind this model   is the  reduction of the  plastic flow problem to a computationally  effective integer-valued \emph{discrete automaton}.  Despite the simplicity of the ensuing dynamical system,   one    can  account in this way  for both short-range and long-range  elastic  interactions, including dislocation nucleation and immobilization.   It also allows one  to     accumulate sufficient statistics,  since one can deal in this way  with  millions of meso-scopic elements and tens of thousands of dislocations.  
  
The 2D version of this model was first introduced in \cite{salman2011,salman2012}, and here  we simply recall  its main characteristics following \cite{zhang2020}. The model assumes that the  displacement field is scalar and that the flow is of  single-slip nature. Hence, forest dislocations cannot be considered directly. However, we recall that   the plastic flow of sufficiently small micro-pillars  is mainly single-slip independently of the underlying crystal symmetry. Even in the case of multi-slip orientation, due to a limited number of available dislocation sources within the confined volume,  the first activated slip  plane dominates and prevents other slip planes from getting involved. In this situation,  the usual  frustration leading to hardening can be avoided considering the absence of dislocation cross-slip and easy annihilation at a free surface.  While any adequate crystal plasticity model would effectively reduce to our constrained single-slip theory in a  sufficiently small system, it should, of course,  allow for multi-slip flow  to take over at larger  sample sizes.  In fact,  a fully tensorial 2D model  has been recently proposed, which allows the modelling of different crystal symmetries and multislip configurations \cite{baggio2019,Salman2019-no}, however at a much higher computational cost.   

In the framework of the scalar model we essentially imply that the sample is oriented for a single slip along the  only available slip direction. The crystal is modeled as an $N\times N$ square  lattice with  the meso-scopic  spacing  normalized to unity.  The deformation of the crystal is given by the displacements of the vertices of the mesoscopic elements, $\vec{u}_{i,j}=(u^x_{i,j},u^y_{i,j})$, where $i,j=1,2,\dots,N$. 

In view of the  single slip assumption we  can set $u^y_{i,j}\equiv0$. We can then    introduce the  notation  $u_{i,j} \equiv u^x_{i,j}$. In the presence of a  kinematic constraint   the strain tensor can be reduced to  two fields: a longitudinal  strain, 
 $\zeta_{i,j} =
u_{i+1,j}-u_{i,j},$  
which is a  linear, non-order parameter variable,  and a shear strain 
 $\xi_{i,j} =u_{i,j+1}-u_{i,j},$ 
which is a nonlinear,  order parameter type variable,  given that  plastic slip originates from multi-well  nature of lattice potential. 

We write  the dimensionless  energy of the system in the form \cite{salman2011}
$
\Phi=\sum_{i,j} f(\zeta_{i,j},\xi_{i,j}),
$
where 
$
f(\zeta,\xi)=( K/2)\zeta^2+f_0(\xi)
$
is the energy of a single (meso-scopic) element. To account for the lattice periodicity we assume that  $f_0(\xi)=f_0(\xi+n),$ where $n \in \mathbb{Z}$ is an integer-valued   slip. Moreover, for analytical transparency we assume that the \emph{periodic} energy density $f_0$ is   piece-wise quadratic  
$
f_0 (\xi_{i,j}) =   (1/2)(\xi_{i,j}-d_{i,j}(\xi))^2 
$.
Here the plastic slip $d$   is represented  by an  integer nearest to $\xi$ so that  
$
   d_{i,j}(\xi)=\lceil{\xi_{i,j}}\rceil 
$
.  
    The obtained model depends on a  single  dimensionless    parameter   $K$ which  mimics the  ratios of elastic constants $ (C_{11}-C_{12})/(4C_{44})$ or $C_{11}/C_{66}$.  It  describes the  coupling between  mesoscopic elements that carry different  values of   $\xi$. In the  limits  $K \to 0,\infty$  we obtain  solvable 1D   models with mean field type interaction \cite{puglisi2005, salman2012}.  At $K \neq 0$  the  model reproduces Eshelby-type propagator and therefore captures crucial effects of long range interactions   induced by elastic compatibility, see more about this below. In our numerical experiments we assumed that $K=  2$ which represents  a typical value for  metallic crystals.
 
The model can be reduced to a discrete  automaton  because  the elastic   problem  $
\partial \Phi/\partial u_{i,j}=0
$
 can be solved analytically  if the  integer-valued field $d$ is known \cite{salman2011}.  The associated equilibrium equations in the bulk, written in terms of the displacement field $u_{i,j}$, read  
\begin{multline}
 K {}(u_{i+1,j} + u_{i-1,j} - 2u_{i,j}) + (u_{i,j+1} + u_{i,j-1} 
- 2u_{i,j}) 
  - (d_{i,j} - d_{i,j-1}) =0.
\end{multline}
The whole system can  be  written in  matrix form 
$\textbf{M} u =b,
$
where $\textbf{M}$ is a pentadiagonal matrix and $b$ is a vector of size $N\times N$ incorporating  the boundary conditions and the field $d$. The problem then reduces to a simple matrix inversion.

We assume periodic boundary conditions  in the horizontal direction $u_{1,j} = u_{N+1,j}$.  The hard device type loading will be applied through the boundary condition in the vertical direction $u_{i,N+1} = u_{i,1}+\gamma$, where $\gamma$ is the control parameter. Periodicity is assumed to allow for the fully explicit  inversion of the matrix $\textbf{M}$. Indeed, we can then use the spectral approach based on  the Fourier transform     $
\hat x(\bold q)= N^{-2}\sum_{ab}x_{a,b}\mathrm{e}^{-{i} \bold q \bold r}
$
  with $\bold  r=(a,b)$ and $\bold q=(2\pi k/N,2\pi l/N)$.  In Fourier space the solution of our linear problem is straightforward and we  can   obtain an explicit representation for   the equilibrium  shear strain
 \begin{equation}
 \hat\xi(\bold q) = \gamma\delta(\bold q) + \hat L (\bold q) \hat d(\bold q), 
 \end{equation}
 where we recall that  $\gamma=\langle\xi\rangle$  is  the measure of the imposed affine deformation. Here the  sign-indefinite  Eshelby-type   kernel with $r^{-2}$  far field asymptotics takes the form
 \begin{equation}
\hat L (\bold q)  = \frac{\sin^2(q_y/2)}{K\sin^2(q_x/2) + \sin^2(q_y/2)}.
\label{kernel1}
 \end{equation}
 Its dipolar structure  reflects the \emph{scalar} nature of our model; the more conventional  quadruple structure of the stress propagator is a feature of isotropic  elasticity, while  here we  deal with the extremely anisotropic limit \cite{picard2004,tyukodi2016}.  

Since we now know  how to update the elastic fields, we can  formulate the quasi-static  athermal dynamics in the form of a discrete \emph{automaton} for the integer-valued field $d$. We 
 start with the unloaded ($\gamma=0$) and   dislocation-free  state ($d_{i,j} \equiv 0$). We then advance the loading parameter $\gamma$ and compute (predict) the elastic field $u_{i,j}$ while keeping the field $d_{i,j}$ fixed. 
 The knowledge of the shear strain field $\xi_{i,j}$ allows us to update (correct) the plastic strain field using the relation $d=\floor*{\xi}$; the update takes place when  the boundary of the energy well is reached   by at least one of the mesoscopic elements.  Then an avalanche occurs while  we use synchronous dynamics for the updates of  $d_{i,j}$. We repeat the prediction-correction steps at a given $\gamma$ till the corrections stop changing the field $d_{i,j}$ and the system stabilizes in a new equilibrium state.  As  the stress in this state  is globally below the threshold and we  can start a new search for the  increment of  $\delta\gamma$ that   destabilizes at least  one unit. As soon as
such an element with  $d_{i,j} \neq\floor*{\xi}$ is obtained we apply  our  relaxation protocol again,  initiating another avalanche.  When avalanche finishes, the variation of $\gamma$ resumes. 
 
In eq. (10), two types of quenched disorder can be introduced \cite{zhang2020}. The \textit{nonlocal} disorder field $h$ mimics the effect of elastically incompatible impurities such as solutes. The \textit{local} disorder $g$ can be viewed as resulting from lattice-compatible obstacles with only a local effect on plastic slip such as e.g. locked dislocation multipoles whose long-ranged fields are screened. The energy density accounting for both types of disorder takes the more symmetric form \cite{salman2011,salman2012}:   
 \begin{equation}
 f (\xi_{i,j},\zeta_{i,j} ) = \frac{K}{2}\zeta_{i,j}^2+\frac{1}{2}(\xi_{i,j}-d_{i,j}(\xi))^2-h_{i,j}\zeta_{i,j}-g_{i,j}\xi_{i,j}.
 \end{equation}
Both disorder  fields, $h$ and $g$,  can be assumed as   drawn independently in each lattice cell from  Gaussian distributions   
$
p_{s}(r)= (2\pi\delta_s^2)^{-1/2}\exp{(- r^2/(2\delta_s^2))},
$
 where $s=(g,h)$.  
The   specificity of the disorder $ g_{i,j} $, representing essentially  a residual plastic strain,   is that it can be  simply combined in the energy density with the actual plastic strain $ d_{i,j}$. For instance, to account for $g$ in the Fourier representation of the elastic solution,  it sufficient to  replace the  field $\hat  d(\bold q)$  by the   sum $\hat  g(\bold q)+\hat d\bold(\bold q)$.  We can then write 
\begin{equation} 
 \hat\xi(\bold q) = \gamma\delta(\bold q) + \hat L (\bold q) \left[\hat d(\bold q) +\hat g(\bold q)\right]  + \hat L_h (\bold q)  \hat h(\bold q), 
\label{automaton3}\end{equation}
where 
\begin{equation} 
\hat L_h (\bold q)  = \frac{\sin(q_x/2)\sin(q_y/2)(\cos(\frac{q_x-q_y}{2})-i\sin(\frac{q_x-q_y}{2}))}{K\sin^2(q_x/2) + \sin^2(q_y/2)} 
\label{fk2}
\end{equation} 
is a distorted  Eshelby  propagator  \eqref {kernel1} maintaining, however, its sign-indefiniteness and the decay rate  $1/r^2$. 

\begin{figure}[h]
 \includegraphics[width=0.8\linewidth]{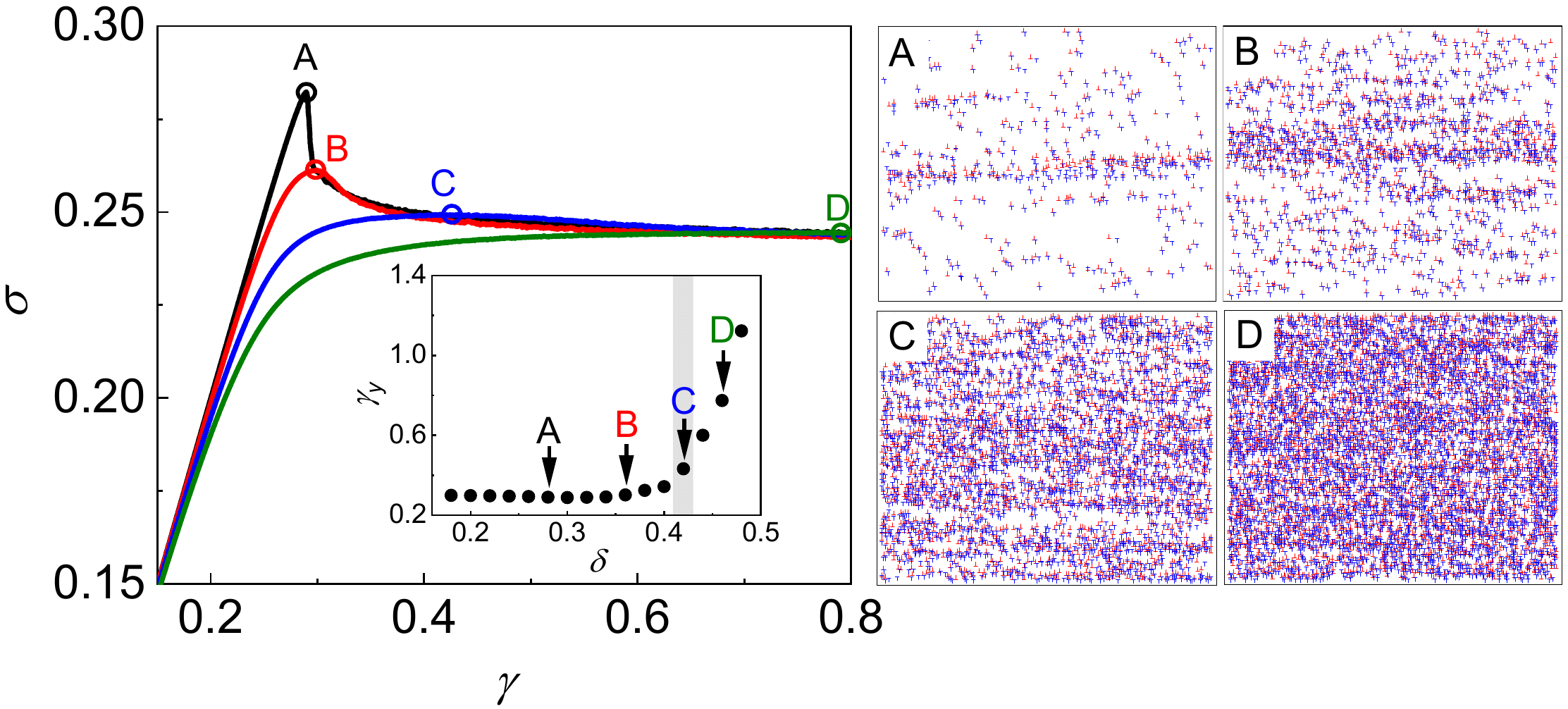}
 \caption{The effect of disorder on average stress-strain curves in simple shear simulations (N=1024). The inset shows the yield strain $\gamma_y$, with the grey strip marking the extended BD transition. Panels A to D show zooms on the corresponding post-yield dislocation configurations.}
 \label{fig:SS-model}
 \end{figure}

Below we first show some  of our simulation results for the case of nonlocal disorder $h$. In all numerical experiments  we considered initially dislocation-free systems ($d \equiv$ 0), and the statistical results  were averaged over at least 100 realizations of the disorder. Figure \ref{fig:SS-model} shows the average mechanical response of our system under simple shear as well as the evolution of the yield strain $\gamma_y$ as a function of the disorder variance $\delta_h=\delta$ under the assumption that  $\delta_g=0$. For weak disorder, $\delta \leq$0.3 (regime A on Fig. \ref{fig:SS-model}), mimicking initially dislocation free, almost pure and small crystals, yielding is brittle-like, with an abrupt stress drop and a strong strain localization along a shear band  which concentrates the dislocations \cite{zhang2020} (panel (A) on Fig. \ref{fig:SS-model}). This regime is reminiscent of the brittle behavior of nanoparticles prepared from solid-state dewetting, i.e. initially dislocation free \cite{mordehai2011,mordehai2018,flanagan2019}, or of our smallest Mo pillars (Fig. \ref{fig:mildtosup}).  

Upon increasing the disorder, the first-order transition eventually terminates at a critical point located around $\delta\simeq$0.42 (regime C), in a way similar to what has been identified in amorphous plasticity \cite{ozawa2018}. In this regime, the plastic slip field $d_{i,j}$ is scale-invariant, characterized by a turbulent-like multifractal pattern \cite{zhang2020}, qualitatively consistent with the spatial fractal pattern of plastic bursts observed from AE in a bulk ice crystal \cite{weiss2003}. At even larger disorder, $\delta \geq$0.5 (regime D), yielding is gradual and the mechanical response is ductile, with both dislocations (Fig. \ref{fig:SS-model} (panel D)) and slip uniformly distributed within the whole crystal \cite{zhang2020}.

\begin{figure}[h]
 \includegraphics[width=0.8\linewidth]{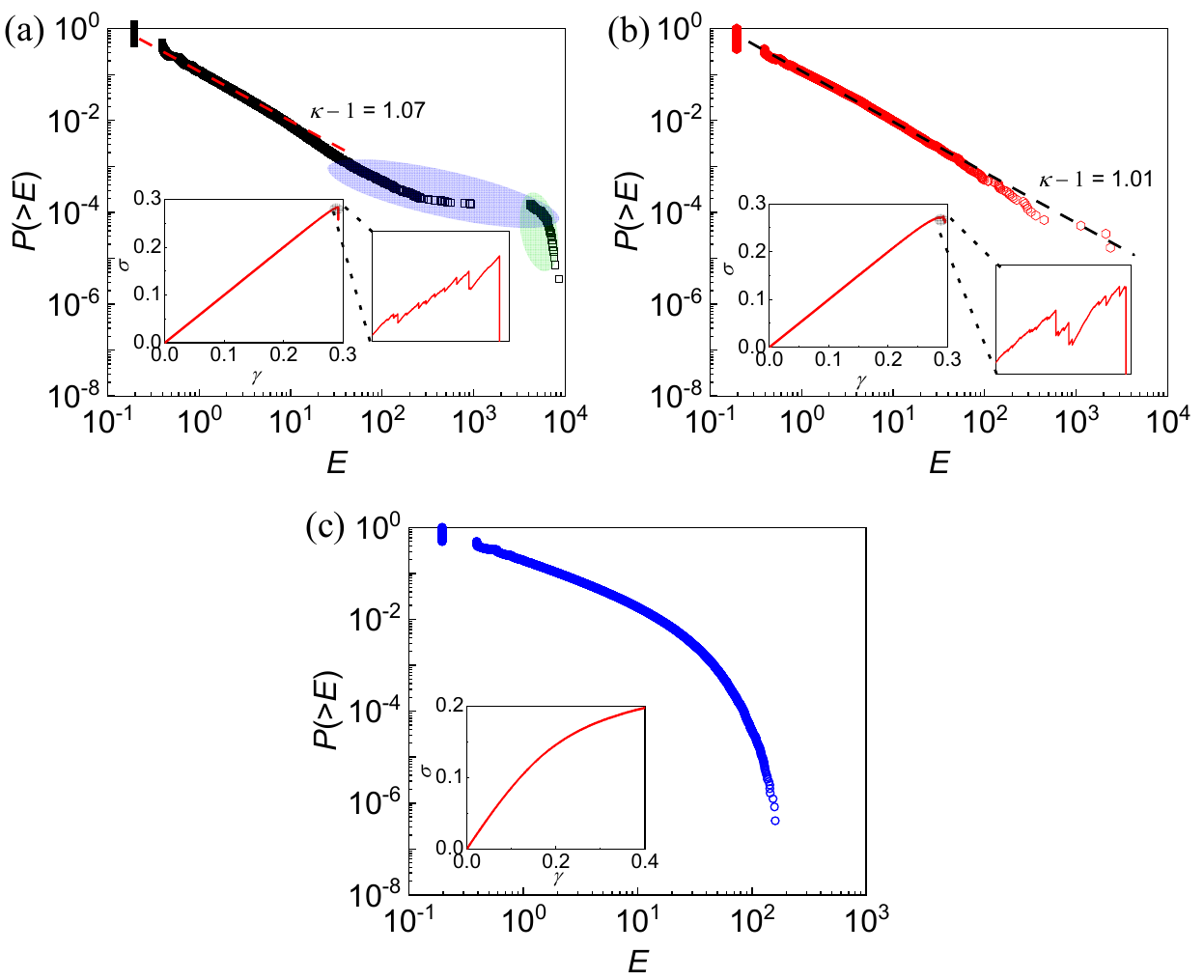}
 \caption{Cumulative probability distributions of preyield avalanche energies at $\delta=$0.28 (a), $\delta=$0.32 (b) and $\delta=$0.7 (c). Averaging was performed over 100 realizations of the disorder. Insets show stress-strain curves fro a particular realization of the disorder.}
 \label{fig:distrib-model}
 \end{figure}

The correspondence with experiments can be also established  in terms of statistics of plastic fluctuations. In our automaton model, the energy $E$ released during an avalanche scales with the cumulative distance covered by all the moving dislocations involved \cite{salman2012,zhang2020}, i.e. with the displacement $X$ as defined above. Hence, computed energy distributions $P(E)$ and experimental distributions $P(X)$ are directly comparable. At small disorder, the model captures the coexistence of dragon-king outliers  with power-law distributed smaller avalanches characterizing a super-critical regime (Fig. \ref{fig:distrib-model}(a)), as observed in our 500 nm Mo pillars (Fig. \ref{fig:mildtosup}(a)). Upon increasing $\delta$, a critical regime emerges (Fig. \ref{fig:distrib-model}(b)), with a power law distribution of avalanches energies and an exponent $\kappa$ consistent with that observed for Mo pillars of intermediate sizes (Fig. \ref{fig:mildtosup}(b)). At even larger disorder, scaling disappears and subcritical statistics are obtained (Fig. \ref{fig:distrib-model}(c)). This general  agreement with observations argues for the robustness of the different regimes identified above, as well as the transitions between them, upon increasing the system size and/or the disorder strength.

In fact, we argue that by varying  the strength of quenched disorder one can  differentiate between sub-micron crystal sizes. Indeed,  instead of $L$ we  should use   a  dimensionless parameter $
R= L/l$ introduced earlier. If we assume that   $l \sim Gb/\sigma_{th}$    identify the threshold    $\sigma_{th}$  with the  pinning (immobilization)  stress, we can recall that the  distinctly brittle regime   would   correspond to  $R  \ll1$, the strongly ductile regime, to $R \gg 1$, while dislocation interaction with obstacles would become relevant  at  $R \sim  1$. 
The threshold $\sigma_{th}$ naturally  depends on the presence of the pinning obstacles and, in general  \cite{zhang2017},  increases with  the variance of  quenched disorder imitating such obstacles. More specifically,  the decrease of  $\sigma_{th}$  can be achieved  by making the disorder more narrow which can be viewed as the way to eliminate  particularly strong obstacles. In this way, instead of increasing  $L$ we  can  decrease  $l$,  which should  be as effective in moving from the brittle  regime, where  $R  \ll1$, to the ductile regime,  where $R \gg 1$. In other words,  instead of exploring directly the dominance of surface effects  one can   exploit  the indirect effect  that  in smaller systems there are fewer strong obstacles that can serve, for instance,  as dislocation nucleation sites because the  existing ones are   compromised or even disabled by their closeness to the surfaces.
 
It  has to be mentioned, however,  that our association of the variance of disorder with crystal size is exclusively targeting systems without bulk criticality, as in the case of Mo crystals.  One can, in principle, manufacture  small crystals with strong (dense) quenched disorder \cite{zhang2017} or grow almost pure large crystals with very weak (sparse) quenched disorder \cite{weiss2019}.  In general, both quenched disorder and  the crystal size  would  affect  brittleness, even though to grow almost defect free crystals (without solutes, precipitates and  dislocations),  is almost impossible except in case of extremely  small sizes (nano-particles).

In  fig. \ref{fig:distrib-model} we showed  stress-\textit{integrated} distributions of plastic fluctuations collected over the entire loading. However, a detailed interpretation of the nature of these fluctuations generally requires an analysis of stress-\textit{resolved} distributions. As an example, a stress-tuned criticality (e.g. depinning) would be characterized by $P(E) \sim E^{-\tau}f(E/E_c)$, where $f(x)$ rapidly vanishes for $x>1$ and $E_c$ is an \textit{upper} cut-off that diverges at a critical stress $\sigma_c$, such that $P(E) \sim E^{-\tau}$ only at the critical point $\sigma=\sigma_c$. Note that in this case the stress-integrated exponent $\kappa$ differs from the stress-tuned exponent $\tau$. It has been argued that the plasticity of micropillars could belong to such stress-tuned criticality \cite{friedman2012,uhl2015}. This interpretation is however disputed \cite{ispanovity2014}, while the analysis of stress-resolved distributions might be difficult owing to a lack of statistics. AE data collected on bulk samples furnish larger catalogs that instead argue against tuned criticality \cite{weiss1997,miguel2001,richeton2006}, at least in HCP materials. 

\begin{figure}[h]
 \includegraphics[width=0.7\linewidth]{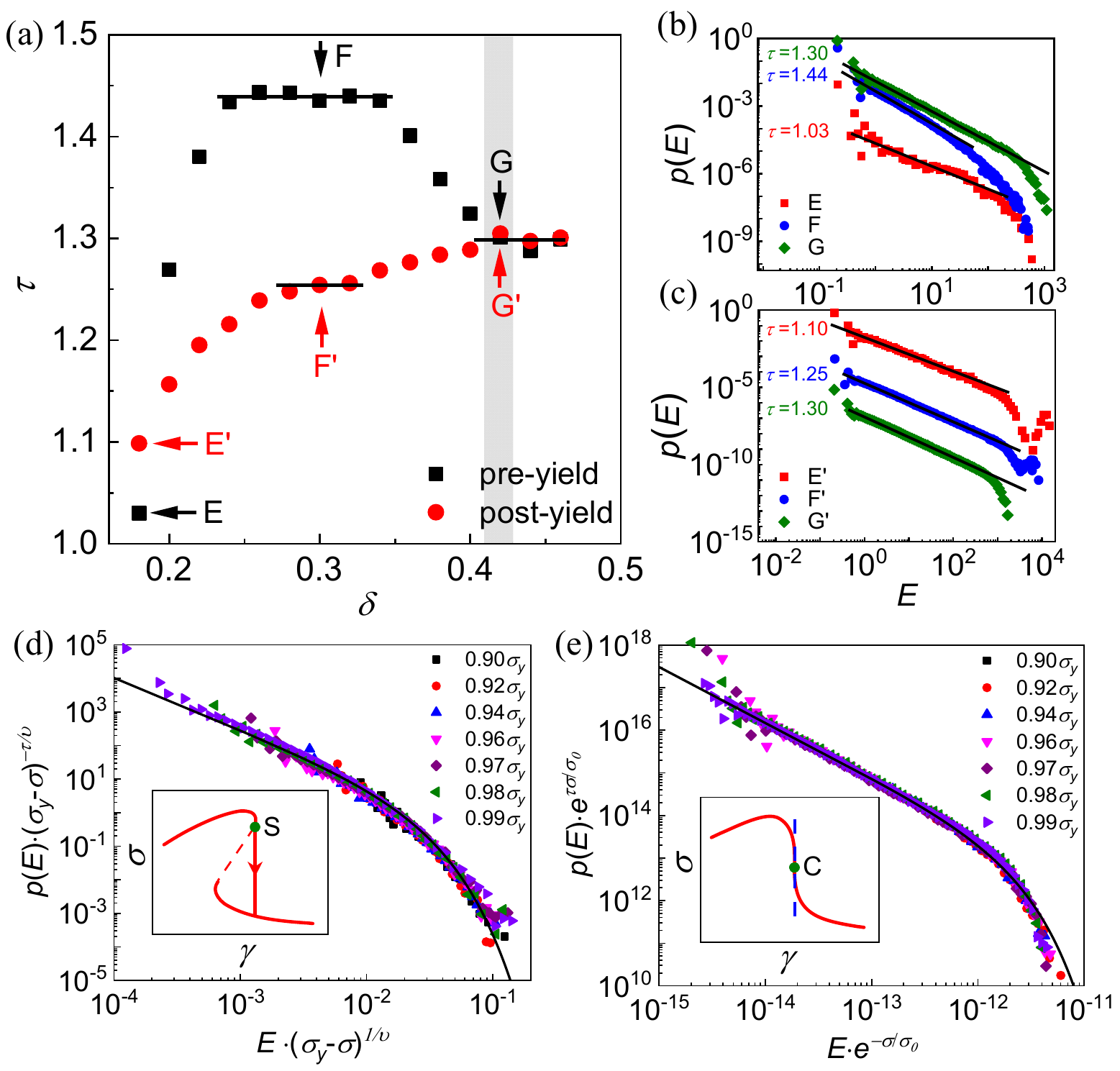}
 \caption{(a) Disorder dependence of the stress-resolved scaling exponent $\tau$ for immediately pre- and post-yield situations. The gray strip marks the schematically the extended BD transition. (b,c) Corresponding avalanche energy distributions. The scaling collapse of the preyield distributions are shown (d) for $\delta=$0.30 (tuned spinodal criticality) and (e) $\delta=$0.46 (BD criticality).}
 \label{fig:exponents-model}
 \end{figure}

Taking advantage of the  low numerical cost of our simulations, we performed a detailed analysis of stress-resolved distributions for the released energies $E$ in our scalar model on the basis of extended statistics. Figure \ref{fig:exponents-model} shows the evolution of immediately pre- and post-yield exponents, along with some examples of corresponding distributions. From these results as well as additional analyses detailed elsewhere \cite{zhang2020}, different types of critical behavior can be identified as a function of disorder strength $\delta$ mimicking also the system size ( see above).

At very small disorder ($\delta \simeq 0.2$), supercriticality and strong brittleness is characterized by small and similar pre- and post-yield exponents, $\tau \simeq 1$. This allows one to draw an analogy with marginal stability of spin glasses \cite{franz2017}. In this case, homogeneously nucleated dislocations self-organize under the influence of long-ranged elastic forces and the system undergoes a transition from a stable (elastic) to a marginally stable (glassy) state.

Over an intermediate disorder range ($0.25<\delta<0.35$), a gap opening is observed between the pre- and post-yield exponents, and a characteristic peak is still observed in the post-yield distribution (Fig. \ref{fig:exponents-model}(c)). A scaling collapse analysis reveals a \textit{tuned} spinodal criticality in this regime, with an upper cut-off $E_c$ diverging as approaching the yield stress as $E_c \sim (\sigma_y - \sigma)^{-1/\nu}$, however with an exponent $1/\nu \simeq$1.6 different from the mean-field depinning prediction $1/\nu =$2 \cite{dahmen2009,salje2014} (Fig. \ref{fig:exponents-model}(d)). This suggest that tuned-criticality could be indeed relevant for the plasticity at small system sizes \cite{friedman2012}. We recall however that our modelled systems are initially dislocation-free, which is hardly the case in micropillar experiments, at least for FCC light materials (see section 2.3). When performing cyclic loading at these levels of disorder with our model, the post-yield stress drop as well as the associated super-critical avalanches disappear from the first loading reversal, i.e. the distributions becomes critical \cite{zhang2020}. This strongly suggests that spinodal tuned-criticality is suppressed when reloading a dislocation-rich system. In other words, a potentially brittle nanocrystal could be "trained" to become more ductile from gentle cyclic loading, with potential applications in nano-engineering. 

Note that the spinodal critical regime is only observed over a limited range of disorder. At larger disorder ($\delta \sim 0.42-0.46$), the BD transition takes place, with the post-yield stress drop as well the characteristic peak in the distribution disappearing, and pre- and post-yield collapsing (Fig. \ref{fig:exponents-model}(a)). This second-order BD criticality is associated with a cut-off following a different asymptotics, $E_c \sim exp(\sigma/\sigma_0)$, where $\sigma_0$ is a constant (Fig. \ref{fig:exponents-model}(e)), meaning that criticality is not stress-tuned in this case \cite{ispanovity2014}. Finally, upon increasing further the level of disorder beyond $\delta \simeq$0.5, "hardening" takes place almost from the onset of loading, plastic activity becomes homogeneous and uncorrelated, and scaling is getting lost.

To summarize, our  numerical studies, consistently with the experimental observations reported above, reveal an extremely rich repertoire of plastic behaviors. An evolution from a typically brittle behavior (though without cracks) to a mostly ductile response can be conceptualized as a complex three-stage crossover: a spin-glass-type marginality encountered for very small, almost disorder-free crystals, transitions to a spinodal stress-tuned criticality at an intermediate level of disorder, then followed by a second-order BD transition at larger disorder (and/or size), to finally a lack of scaling and a fully ductile behavior at very large disorder/scale. This scenario shows some similarity with what has been recently proposed for amorphous plasticity \cite{ozawa2018,popovic2018}, although crystalline plasticity appears even more intricate. In this framework, scaling laws and exponents are non-universal.

\begin{figure}[!htbp]
	\centering	
	\includegraphics[scale=.65]{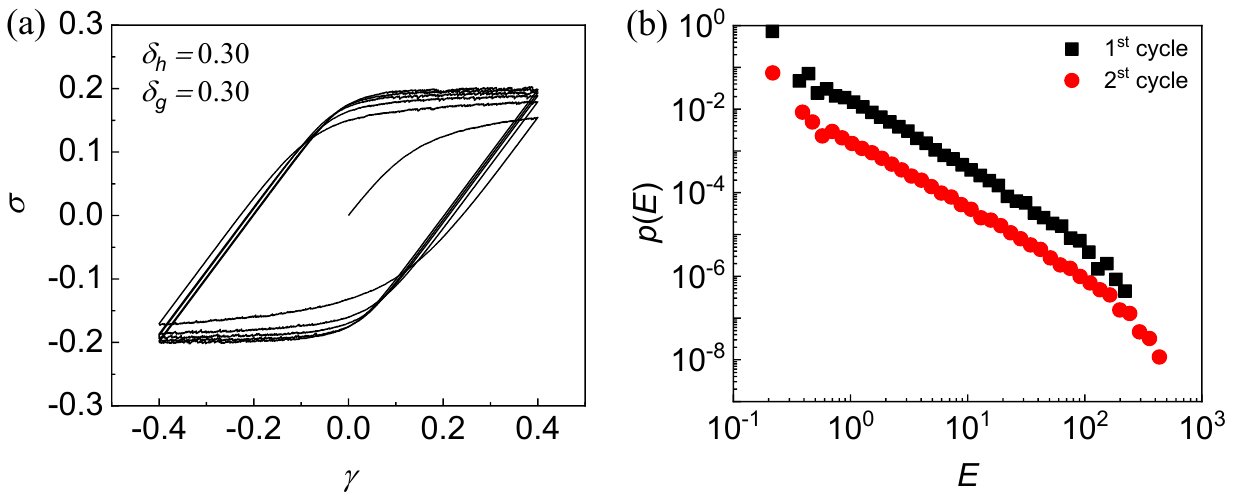}	
	\caption{(a) Strain-stress curves for the crystals subjected to six loading unloading cycles; (b) Avalanche distributions of cyclically loaded crystals for the first and the second cycles;  the first cycle is understood as  the monotone loading path.  Here $\delta_h=\delta=0.30$, $\delta_g=0.30$.
		\label{cyclic_distribution2}}
\end{figure}

We now briefly illustrate  the interplay between our two types of disorder, 'local' and 'nonlocal'.   To avoid the dependence on the initial preparation we have now choose  the setting of cyclic loading. Our numerical experiments, summarized in Fig. \ref{cyclic_distribution2}(a), show  that when a weak 'local' disorder $\delta_g=0.3$ is combined with a weak 'nonlocal'  disorder $\delta_h=0.3$,  the overall mechanical response is  ductile. The  initial softening behavior, observed in crystals with $\delta_g=0$,  is replaced by the more conventional  hardening behavior. At large strains   the stress response  shows  a robust yielding plateau independently of the configuration of disorder. The overall response  is  reminiscent of the  classical   strain-hardening behavior  exhibited   by \emph{bulk} FCC and BCC materials \cite{suresh1998}.    

From  Fig. \ref{cyclic_distribution2}(b)  we see that even a weak 'local' disorder is  sufficient to  suppress  super-criticality and to completely eliminate system-size events.  This observation agrees with the idea that such disorder generates local inhomogeneities  which inhibit global  response. However, the increase of the cut-off size in the second cycle  suggests  that  a correlated behavior,  reminiscent of disorder-induced self-organization towards classical criticality in RFIM (Random Field Ising Model)\cite{dahmen1996,da2020}, can still take place.

\begin{figure}[!htbp]
\includegraphics[scale=0.26]{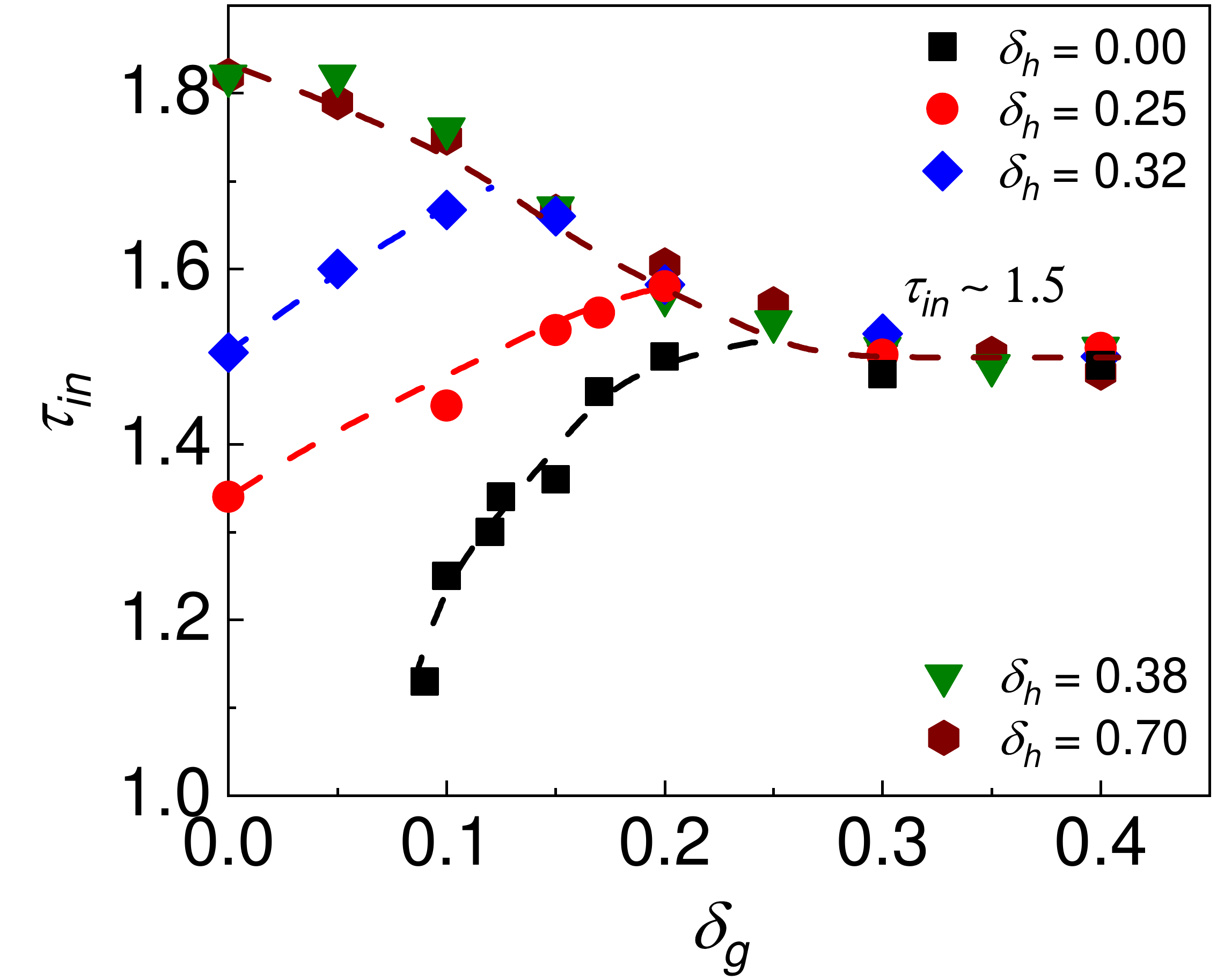}
\caption{   Effect of the 'local' disorder $\delta_g$ on the (integrated) scaling exponent $\tau_{in}$ for  the case of cyclic loading.
\label{fig811}   }
\end{figure}

In Fig. \ref{fig811} we show how the different  configurations of 'local' and  'nonlocal' disorder strengths  affect the cycle-averaged  (integrated) scaling exponents $\tau_{in}$. When the   'local' disorder is weak, we recover the after-yield behavior studied above. At stronger 'local' disorder, the dependence of the exponent $\tau_{in}$ on the 'nonlocal' disorder progressively diminishes. Given that the statistics is mostly  acquired during   hardening-free yield, see  Fig. \ref{cyclic_distribution2}, one can expect the stress resolved value of the exponent $\tau$ to be similar to the aggregate value  $\tau_{in}$ \cite{durin2006}.  In this case the  obtained exponent value  suggests  mean field criticality \cite{dahmen1996,da2020}. In other words, the abundance  of 'local' disorder apparently   trivializes  the scaling picture, erasing the non-universality and promoting a universal response of the athermally driven infinite dimensional RFIM dominating the response of amorphous solids \cite{ozawa2018,ozawa2020,bhaumik2019,franz2020}.

However, the overall  agreement between our  automaton  model, which is based on some crude assumptions, and experiments,  is  incomplete. Thus, in experiments,  we encounter  distributions of avalanche sizes mixing a power law tail with a \textit{lower} cut-off below which fluctuations are mild. This wild-to-mild coexistence  and its unique signatures, such as the relationship between $\kappa$  and the wildness $W$ (Fig. \ref{fig:universal curve}(c)), are not recovered within this oversimplified model. The problem is most probably in the the single-slip assumption that prevents the emergence of metastable dislocation  patterns. Another shortcoming  is  that disorder is prescribed as a single-scale field so that,  by construction, fluctuations at smaller scales are compromised.

\subsection{Mean-field model}

 A simple  mean field  model  can be used to rationalize  at least some elements of the observed bigger picture  in terms of  macroscopic parameters. In particular, it will allow us to  explain the coexistence of mild and wild fluctuations in crystalline plasticity. 
 
Suppose  that the stress resolved evolution of the spatially averaged density of \textit{mobile}  dislocations  $\rho$  is described  by a stochastic kinetic equation   \cite{weiss2015}
\begin{equation}
 \rho^{-1}d\rho/ d\gamma= a\rho^{-1}-c+\sqrt{2D}\eta(\gamma),                                         
 \label{stoch}
\end{equation}
where  the local shear strain $\gamma$ serves as a time-like parameter,  $c \geq 0$ characterizes the  rate of dislocation immobilization and   the  
temperature-like parameter $D$ represents the intensity of the multiplicative mechanical noise with  $\langle \eta(\gamma) \rangle=0$ and $ \langle \eta(\gamma_1 ),\eta(\gamma_2) \rangle=\delta (\gamma_1  -\gamma_2)$.  In view of the constant stress assumption, the route towards yielding in a stress-tuned regime cannot be described in this way,   however  this model can be used to  rationalize the universal dependence of wildness on both material characteristics and  sample size. 

While the deterministic part of the model (the first two terms on the RHS of  \eqref{stoch}) is quite conventional, in this simplified framework the long-ranged stochastic interactions are described through a multiplicative noise. The level of noise $D$ quantifies the intensity of mechanical fluctuations experienced by a meso-volume due to interactions with the rest of the system. A concept of 'mechanical temperature'  has been also used in the modelling of athermal amorphous plasticity \cite{nicolas2014}. Considering the Orowan's relation $d\gamma=\rho bvdt$, and assuming a constant dislocation velocity $v$ under constant stress, we can link the fluctuations of $\rho$ with the experimentally measured strain (or slip $X$) fluctuations.  

 The   stationary probability distribution in  \eqref{stoch} is 
$
 p_s (\rho)   \sim  e^{-a/(D \rho)}\rho^{-\alpha}
$
with the exponent $\alpha=1+c/D$.  This  is exactly the same expression as  our empirical  eq. (1) with $\kappa-1=c/D$ and $X_0\sim a/D$ \cite{weiss2015,zhang2017}. Consequently, the wildness $W$ is given by expression (3). Therefore, though oversimplified, our mean field  model predicts a relationship between the exponent $\kappa$ and the wildness $W$, which are linked through the characteristic size $X_0$ which represents a material constant.  We reiterate that the existence  of  such  universal  relation between the structure of the power law tail of avalanche distribution  and  the wildness parameter    $W$ is  in full agreement with the observed data  for  a large number of pure materials and alloys (Fig. \ref{fig:universal curve}(c)).

In the framework of our automaton  model we can interpret  $\rho$ as the density of mobile dislocation during an avalanche at a given value of the loading $\gamma$.  We can then write $\rho(\gamma)=n(\gamma)/N^2$,   where $n(\gamma)$ is the number of dislocations moved during  an  avalanche.  Our numerical experiments suggest that the avalanche energy $E$ is a disorder independent linear function of the total distance traveled by mobile dislocations during an avalanche $\bar  l$ and that  $\bar l \sim n$. Therefore  $E \sim \rho$  and   we can conclude  that  the exponent $\alpha$ in the mean field   model  should be indeed the same as  the exponent $\kappa$ in the automaton model. 
 
For  single slip  pure  nano-crystals with weak disorder, dislocation immobilization  can be neglected,   so  $c/D \ll 1$, and   the stochastic evolution of  $\rho$ governed by  \eqref{stoch} reduces  in this case to  a geometric Brownian motion with $\alpha\sim  1$. In the automaton model we  observe in the  low-disorder  limit dislocation self-organization,  governed exclusively by elastic long-range elastic interactions \cite{ispanovity2014,weiss2019},  and recover the same value of the exponent $ \kappa \sim  1$.   With increasing disorder, the immobilization rate $c$ should increase leading to a higher value of $\kappa$,  which is in qualitative agreement with our numerical experiments.
 
 The  crossover from  $D$-dominated brittle regimes  ($c<D$ with the stochastic term in  \eqref{stoch} controlling the dynamics) to  $c$-dominated ductile regimes ($c>D$ with the deterministic term in  \eqref{stoch} controlling the dynamics) can be expected where the  mechanical agitation is balanced by dislocation self-locking  ($c \sim  D$). Using the relation $\alpha=\kappa$,  we can now link $c/D$ and  $R=L/l$. The effective temperature $D$ should depend only weakly  on  the system size $L$.  It is defined instead  by the  locking strength of defects, which means that it increases  with $ l$.  At the same time, it is clear that  the rate of dislocation reactions (in particular our parameter $c$  controlling immobilization)   increases   with $L $ \cite{zhang2017}. Therefore,  in either very small  and/or  very weakly disordered samples   $c<D$. Conversely,  in either bigger or more disordered  samples one  can  expect to reach the  ductile phase  where $c>D$.  We recall that all these trends were observed in our automaton model. 

We can go a little further in the interpretation of the model parameters focusing now on the role of the parameter $a$. In fact, the model has two characteristic densities $\rho_c=a/c$ and $\rho_D=a/D$ or, in other words, two characteristics length scales $l_c=1/\sqrt{\rho_c}$ and $l_D=1/\sqrt{\rho_D}$. The scale ratio $r=l_c/l_D=\sqrt{c/D}=\sqrt{\kappa-1}$ is then the main dimensionless  parameter of the mean field model and it should then  control the wildness $W=W(\kappa)=W(r)$, see eq. (3). On the other hand, we argued, and showed experimentally (Fig. \ref{fig:universal curve}(b)) that $W$ is controlled by the ratio $R=L/l$. By comparing the functions $W(r)$ and $W(R)$ we  find that $R\sim r^2$ because the relation $\kappa-1=c/D \sim L/l=R$ has been verified experimentally \cite{zhang2017}. This suggests $c \sim L$ and $D \sim l$, i.e., for a given material, $c$  expresses an external size effect, while $D$ accounts for an internal scale effect \cite{zhang2017}. 

In particular, in HCP pure materials, such as Ice or Mg, single-slip plasticity and the absence of forest hardening implies a negligible immobilization of dislocation pairs, i.e. a small $c$ value. In addition, for the reasons already discussed in section 2.4, the pinning strength $\tau_{pin}$ there is small, which we can now interpret as a large mechanical temperature $D \sim l \sim 1/\tau_{pin}$ \cite{zhang2017}. This combination gives a small $\kappa$, close to 1, and a large wildness, as observed (Fig. \ref{fig:universal curve}(c)).  On the reverse, for hardened and/or alloyed large FCC metallic  samples, we expect a small value of $D$ as well as an enhanced immobilization term   resulting from the ubiquity of locks and junctions. All this means  large $\kappa$ and a small $W$.  

\section{Conclusions}

Dynamical fluctuations have been for a long time  overlooked in the analysis of crystalline plasticity, with the exception of  the seminal work of Becker and Orowan \cite{becker1932} and the seemingly unrelated   work on  dynamic strain aging \cite{kubin2002}. An implicit  assumption has been that such fluctuations average out when considering "large enough" spatial and temporal scales. The situation has changed over recent years as a consequence of progressive miniaturization of systems and devices.  As a result, the mechanical properties of metallic materials at micro- to nano-scales became a major concern in the material science community. The classical metallurgical practices have been developed and refined for a very long time to optimize different properties, such as strength, formability, resistance to fatigue, ect. for samples at macroscopic scales. Similar questions now arise in microscopic  metallurgy  dealing with sub-µm scales.

In this review we have presented  an outlook on plastic fluctuations in pure materials and alloys with quenched disorder. It reveals a rich and intricate landscape of behaviors and scaling properties that defies conventional phenomenological approaches and calls for a paradigm change  that will open materials science and  metallurgy to the  powerful methods and techniques of nonequilibrium statistical physics. 

In HCP materials, characterized by a low lattice friction and a strong plastic anisotropy, dislocation avalanches are detectable even on macroscopic  bulk scales. They are   power law distributed in size and energy which  suggests  critical dynamics of the type characterizing developed  turbulence. However, at macroscopic  scales, such wild fluctuations are nearly undetectable in most of the situations  of interest to  classical metallurgy.  This is particularly true for  bulk FCC and BCC metals and their alloys, especially once strain hardening takes place. This   explains the lack of interest to this topic until very recent times when ultra-small structures started to dominate industrial applications.  

 If the new  "smaller is stronger" size effect emerging at these scales might appear beneficial at first glance,  it has been shown to be  corrupted by wild plastic fluctuations, possibly reaching the system size and leading to a brittle-like behavior. Consequently, despite the extremely high strength achievable at ultra-small scales, the ensuing plasticity turned out to be  uncontrollable due to the stochastic nature of strain bursts reminiscent of macroscopic earthquakes.  The disastrous dislocation avalanches, playing the role earthquakes at these scales,  may poison the forming processes and compromise the load-carrying capacity in various engineering/industrial processes dealing with sub-µm parts, particularly in nanoimprint lithography \cite{gao2014} or for the shaping of MEMS \cite{hu2016}. Therefore, an urgent  challenge facing today's  metallurgy at sub-µm scales is to reduce the “wildness” of the associated fluctuations, while keeping or improving other properties, such as strength. 

In this review  we focused on  the important fact that the transition from a mild to a wild plasticity is controlled by a dimensionless ratio of length scales which we denoted by $R=L/l$, where $L$ and $l$ are external and internal scales, respectively. In FCC pure materials, the internal scale $l$ is linked to a dislocation mean free path, hence to dislocation patterning and hardening, while lattice friction might play a significant role in shaping the value of $l$ in BCC materials at low temperatures. The introduction of $R$ as controlling parameter  also suggests that plastic intermittency can be reduced or  shifted towards smaller system sizes, by introducing quenched disorder through  alloying and other similar means. Our experimental results fully support the feasibility of such a  "dirtier is milder" metallurgical strategy.  

While these first steps in harnessing plastic fluctuations at ultra-small scales have been made, many key  challenges persist. For instance:

(i) In case of HCP materials, the possibility to tame wild fluctuations at bulk scales, by introducing tailored disorder, remains to be explored.

(ii) Finding the effect of  lattice friction and thermally activated processes below the athermal temperature on plastic fluctuations in BCC materials requires  a systematic analysis.

(iii)  In weakly disordered and dislocation-free or strongly starved FCC crystals, we  identified a transition from a spin-glass type marginality to a spinodal stress-tuned criticality. Both regimes are associated with deleterious system-spanning instabilities and a brittle-like behavior. It is still unclear how such brittleness can be  controlled  without introducing stronger doping.

(iv) At small scales,  stochasticity results in an increasing variability of “global” mechanical characteristics  including strength \cite{rinaldi2008,zaiser2013} and  hardening coefficient (e.g. Fig. \ref{fig:SH}). This large scatter can be considered as another deleterious effect  and this problem  was  recently addressed in case of nano-indentation \cite{derlet2016}. For instance,  as in fracture of disordered materials (e.g. \cite{weiss2014}), the analysis of finite-size effects on strength \textit{variability} is of crucial importance.  

(v)  The conceptual stochastic models of plasticity, discussed in section 5, remains too schematic to adequately account for geometry of real systems and the complexity of the associated loading protocols. In other words, realistic problems  are still outside the realm of  numerical modeling by stochastic differential equations.  DDD simulations have been extensively used to model the associated  systems but the reach of this   approach remains limited  because of the ever-increasing set of ad-hoc rules required for its implementation. An alternative may be linked with  embedding of stochastic rheological closure relations within (so far) deterministic finite-element (FE) codes. This could be done by introducing explicitly fluctuations, instabilities and scatter into the standard numerical codes used in engineering applications. 

\section{Acknowledgements}  This work was supported by  
 French-Chinese ANR-NSFC grant (ANR-19-CE08-0010-01 and  51761135031).
  P.Z.  acknowledges additional support from China Scholarship Council and China Postdoctoral Science Foundation (grant 2019M653595).

% The ext command determines the bibliography style. Please do not
% change this.
\bibliographystyle{crplain}

%This calls all references from the .bib
\nocite{*}

%  This inserts the bib file
%\bibliography{Biblio}

\def\bysame{\leavevmode ---------\thinspace}
\makeatletter\if@francais\providecommand{\og}{<<~}\providecommand{\fg}{~>>}
\else\gdef\og{``}\gdef\fg{''}\fi\makeatother
\def\cdrandname{\&}
\providecommand\cdrnumero{no.~}
\providecommand{\cdredsname}{eds.}
\providecommand{\cdredname}{ed.}
\providecommand{\cdrchapname}{chap.}
\providecommand{\cdrmastersthesisname}{Memoir}
\providecommand{\cdrphdthesisname}{PhD Thesis}

\end{document}